\documentclass[12pt]{article}
\usepackage[margin=0.9in]{geometry}

\usepackage{amsmath}
\usepackage{amssymb}
\usepackage{bm}
\usepackage{graphicx}
\usepackage{authblk}
\usepackage{xcolor}
\usepackage{cancel}
\usepackage{amsthm}
\usepackage{hyperref}
\usepackage{subcaption}
\usepackage{lineno}

\newtheorem{theorem}{Theorem}
\newtheorem{corollary}{Corollary}

\newtheorem{lemma}{Lemma}
\newtheorem{proposition}{Proposition}

\theoremstyle{definition}
\newtheorem{definition}{Definition}
\newtheorem{remark}{Remark}
\newtheorem{example}{Example}

\begin{document}

\title{Lecture notes on classical and quantum non-Markovianity} 
\author{Graeme Pleasance\footnote{gpleasance1@gmail.com}}

\affil{National Institute for Theoretical and Computational Sciences (NITheCS), Stellenbosch, 7600, South Africa}
\affil{Department of Physics, Stellenbosch University, Stellenbosch 7600, South Africa}

\date{\today}

\maketitle

\begin{abstract}
    The study of non-Markovian quantum processes has attracted significant interest in recent decades, giving rise to several competing notions of quantum non-Markovianity.
    These notes serve as an introduction to the topic for graduate students familiar with quantum mechanics and probability theory. 
    Owing to the vastness of the literature, we focus on two prominent characterizations of quantum Markovianity based on the divisibility of quantum channels and monotonically decreasing state distinguishability.
    The correspondence between classical concepts (stochastic matrices, Chapman-Kolmogorov equation) and their quantum analogs (dynamical maps, CP-divisibility) is emphasized throughout.

    
\end{abstract}

\section{Introduction}

These notes review some introductory concepts underpinning the characterization of non-Markovianity in classical and quantum stochastic processes.
Since the classical definition of Markovianity is incompatible with the statistical foundation of quantum mechanics \cite{Breuer2016},
the question of what is precisely meant by quantum non-Markovianity has been a topic of growing interest in recent decades \cite{Li2019,Breuer2012,Rivas2014}.
The basis on which this question has sought to be addressed is the theory of open quantum systems \cite{BP2002,Rivas2010}, 
which treats noise effects and stochasticity as emerging from the interaction between a quantum system and its environment.
Within this framework, the time evolution of an open quantum system can be described using the dynamical map \cite{Alicki1987,Chruscinski2022}, 
which is constructed by tracing over the environmental part of the unitarily evolved system-environment density matrix. 
This map provides a reduced description of the dynamics insofar that it evolves the density matrix of the open system without 
explicitly tracking system-environment correlations or changes to the environment state.
In this aspect, one may consider two distinct ways of characterizing quantum memory effects \cite{Buscemi2018}: 
the reduced or `intrinsic' approach, which associates non-Markovianity with specific properties of the dynamical map \cite{Breuer2009,Laine2010,Rivas2010b,Vacchini2011,Chruscinski2014};
or the `extrinisic' approach, which associates non-Markovianity with global properties of the system and environment, including system-environment correlations \cite{Pollock2018a, Pollock2018b, Gullo2014, Milz2020,Guarnieri2014}. 
A review of various non-Markovianity criteria relating to these approaches can be found in \cite{Li2018}. 
In these notes, we focus solely 
on intrinsic Markovianity criteria based on CP (completely positive)-divisibility \cite{Rivas2010,Rivas2010b} and monotonically decreasing state distinguishability \cite{Breuer2016,Breuer2009}. 
The material is intended for students with an advanced undergraduate or early postgraduate level knowledge of quantum mechanics and probability theory.

The remainder of these notes are organized as follows. 
In Section \ref{sec:2}, we recap the basics of classical probability theory and stochastic processes, including Markov chains and notion of P-divisible stochastic maps. 
In Section \ref{sec:3}, we review the theory of open quantum systems and develop the groundwork needed to formulate the CP-divisibility and trace distance measures of quantum Markovianity (and therefore non-Markovianity), which are subsequently introduced in Section \ref{sec:4}.
Finally, we explore an application of these measures to the spin-boson model in Section \ref{sec:5}. 
Exercises with accompanying solutions may also be found at the end of the notes.

\section{Classical probability theory}\label{sec:2}

To set the notation, we will first review basic concepts from classical probability theory,
before proceeding to formulate classical Markov processes. 
Further detail on these topics can be found in many of the standard textbooks on probability theory and stochastic processes \cite{Gardiner2009, Kampen2007}.
Likewise, parts of the material on classical Markovianity---and later discussions on quantum definitions of Markovianity---will draw on the work of recent review papers \cite{Breuer2016,Li2019,Rivas2014,Li2018,deVega2017}.

A classical random process is described by a random variable $X:\Omega\rightarrow \mathbb{R}$ on a probability space $(\Omega, \mathcal{F}, P)$.
The three key ingredients of a probability space are: the sample space $\Omega$, 
the $\sigma$-algebra\footnote{The $\sigma$-algebra of $\Omega$ is the family of all subsets of $\Omega$ that includes both $\Omega$ and 
the empty set $\emptyset$, and which is closed under countable unions and under complements.} 
$\mathcal{F}$ of all subsets of $\Omega$, and the 
probability measure $P$.
The sample space represents all possible states that the system may occupy, while $\mathcal{F}$ represents the family of events which may be formally assigned a probability through the mapping
$P:\mathcal{F}\rightarrow [0,1]$. 
In particular, if $B\in\mathcal{B}(\mathbb{R})$, then $P_X(B) = P(X^{-1}(B))$ defines the probability distribution of $X$ from the 
measure $P$, where $X^{-1}$ denotes the pre-image of $B$ in $\mathcal{F}$, and $\mathcal{B}(\mathbb{R})$ denotes the Borel $\sigma$-algebra of all 
open intervals on $\mathbb{R}$.
The measure $P$ must satisfy the properties:  
\begin{enumerate}
    \item Normalization $P(\Omega) = 1$. 
    \item Positivity $0\leq P(F)\leq 1$ for all measurable sets $F\in\mathcal{F}$. 
    \item Additivity $P(\bigcup_i F_i) = \sum_iP(F_i)$ for all $F_i\cap F_j = \emptyset$, $i\neq j$.
\end{enumerate}
For simplicity, we shall restrict ourselves to finite sample spaces $\Omega = \{\omega_1,...,\omega_d\}$ with $d<\infty$.

The definition of a random variable may also be extended to a random vector $X=(X_1,...,X_n)$ in $\mathbb{R}^n$.
In this case, the probability distribution for $X$ is the joint probability distribution $P_n(X_n=x_n,...,X_1=x_1) = P(X^{-1}(B_1\times ...\times B_n))$.
For shorthand we shall use the notation $P_n(x_n,...,x_1):=P_n(X_n=x_n,...,X_1=x_1)$.
It is also possible to define conditional probabilities through Bayes' rule,
\begin{equation}\label{eq:cond_prob}
    P_{n-m|m}(x_n,...,x_{m+1}|x_m,...,x_1) = \frac{P_n(x_n, x_{n-1},...,x_1)}{P_m(x_m,...,x_1)}, \quad n>m,
\end{equation}
which represents the joint distribution of $X$ when the variables $X_1,...,X_m$ are fixed. 
Finally, the probability distribution corresponding to the subset of $n-m$ random variables $(X_{m+1},...,X_n)$ sampled from $X$ is defined as
\begin{equation}
    P_{n-m}(x_n,...,x_{m+1}) = \sum_{x_1,...,x_m}P_n(x_n,...,x_m,...,x_1),
\end{equation} 
called the marginal distribution of $P_n$.

\subsection{Stochastic processes}

A stochastic process is essentially a time dependent random process described by a random variable $X_t$, whose statistical properties varies with time. 
More formally, it is defined as a one-parameter family of random variables $\{X_t : t \in T\}$ on a 
common probability space $(\Omega, \mathcal{F}, P)$, where $T$ is a non-empty index set or interval $T\subset \mathbb{R}$. 
At fixed $t$, the function $\omega \mapsto X_t(\omega)$ behaves like a random variable defined in the same way above. 
On the other hand, when $t$ varies it may do so either discretely or continuously depending on how the system is measured.
In the following we shall focus solely on the discrete case where $T=\{t_1,t_2,...,t_n\}$ and $t_n\geq t_{n-1}\geq ...\geq t_1$.

A trajectory is defined as a single realisation of a stochastic process, i.e., a sequence of values $\{X_t(\omega) : t \in T\}$ corresponding to a fixed sample $\omega \in \Omega$. 
The collection of all possible trajectories (or histories) forms the sample space $\Omega$ of the stochastic process.

Some examples of discrete-time stochastic processes include: 
\begin{itemize}
    \item Flipping a two-sided coin $n$ times. 
    Here, $T = \{0,1,...,n\}$, and the sample space is $\Omega = \{{\rm H},{\rm T}\}^n$, with $\{{\rm H},{\rm T}\}$ representing 
    the outcome of a single coin flip.  
    A trajectory for this process corresponds to any sequence of $n$ flips, e.g., $\omega = ({\rm H,T,T,H,H})$. 
    The structure of the sample space is equivalent to a random process where $n$ coins are flipped simultaneously.
    
    \item Random walk on a $d$-dimensional lattice: $T=\{0,1,...,n\}$ and $\Omega = \mathbb{Z}^d\times ... \times\mathbb{Z}^d$.
\end{itemize}
An important quantity in the study of stochastic processes are the finite joint probability distributions
$P_n(x_{t_n},...,x_{t_1})$, which encode the joint statistics of the sequence of random variables $\{X_{t_1}, X_{t_2}, ..., X_{t_n}\}$.
Since the index $n$ may be varied, this process will be characterized by the hierarchy of all such joint distributions for $n\in\mathbb{N}$.
Additionally, these distributions must obey the so-called Kolmogorov consistency conditions to define a valid stochastic process.
The first of these conditions states that any family of joint distributions must be closed under marginalization. 
The second stipulates that all joint distributions obey permutation symmetry in their time arguments.
Mathematically, these are equivalent to
\begin{align}
  P_{n-1}(x_{t_n}, ..., \cancel{x_{t_j}},..., x_{t_1}) &= \sum_{x_{t_j}}P_n(x_{t_n}, ..., x_{t_j},...,x_{t_1}),  \label{eq:kcc1} \\
  P_n(x_{t_n}, ..., x_{t_1}) &= P_n(x_{\pi(t_n)},...,x_{\pi(t_1)}), \label{eq:kcc2}
\end{align}
where $\pi(t_j)$ denotes a permutation of the argument $t_j$, e.g. $\pi(t_1) = t_2$.
The intuition behind \eqref{eq:kcc1} is that performing a noninvasive and `nonselective' measurement at time $t_j$ (i.e., a measurement where the outcome $x_{t_j}$ is not recorded) should not affect the postierior statistics of $X_t$,
while \eqref{eq:kcc2} infers that the time label $t_j$ plays a benign role in the construction of $P_n(x_{t_n},...,x_{t_1})$:
namely, these can either reference different times over which a single system is measured, or reference multiple subsystems measured at a single time \cite{Li2018}.

\subsubsection{Markov processes}\label{sec:2.1.1}

The conditional probabilities of a stochastic process are defined analogously to \eqref{eq:cond_prob} as 
\begin{equation}\label{eq:st_cond_prob}
    P_{n-m|m}(x_{t_n},...,x_{t_{m+1}}|x_{t_m}, ..., x_{t_1}) = \frac{P_n(x_{t_n}, x_{t_{n-1}}, ..., x_{t_1})}{P_{m}(x_{t_{m}}, ..., x_{t_1})}, \quad n>m.
\end{equation}
For a general stochastic process, the probability of measuring $x_{t_n}$ at time $t_n$
depends on the values $X_t$ takes at previous times. 
In particular, if $X_{t_n}$ depends on all previous states $x_{t_{n-1}}, ..., x_{t_1}$, then the full hierarchy of conditional probabilities \eqref{eq:st_cond_prob}
and initial probability distribution $P_1(x_{t_1})$ is required to provide a complete statistical description of the process. There exist a special class of processes, however, for which only $P_1(x_{t_1})$ and two-point conditional probabilities
are needed, known as Markov processes. 
Formally, they adhere to the following definition.

\begin{definition}[Markov process]\label{def:Markov}
    A stochastic process $\{X_t : t \in T\}$ is said to be Markovian if for all $n\geq1$, the conditional probabilities 
    $P_{1|1}(x_{t_n}|x_{t_{n-1}},...,x_{t_1})$ are independent of the values that $X_t$ takes at times prior to
    $t_{n-1}$:
    \begin{equation}\label{eq:Markov}
        P_{1|n-1}(x_{t_n}|x_{t_{n-1}},...,x_{t_1}) = P_{1|1}(x_{t_n}|x_{t_{n-1}}),\quad n\geq1. 
    \end{equation} 
\end{definition}

\begin{remark}
    A discrete-time stochastic process with the Markov property \eqref{eq:Markov} is also known as a Markov chain \cite{Norris2012}.
\end{remark}

\begin{figure}[t!]
\centering
\includegraphics[scale=0.38]{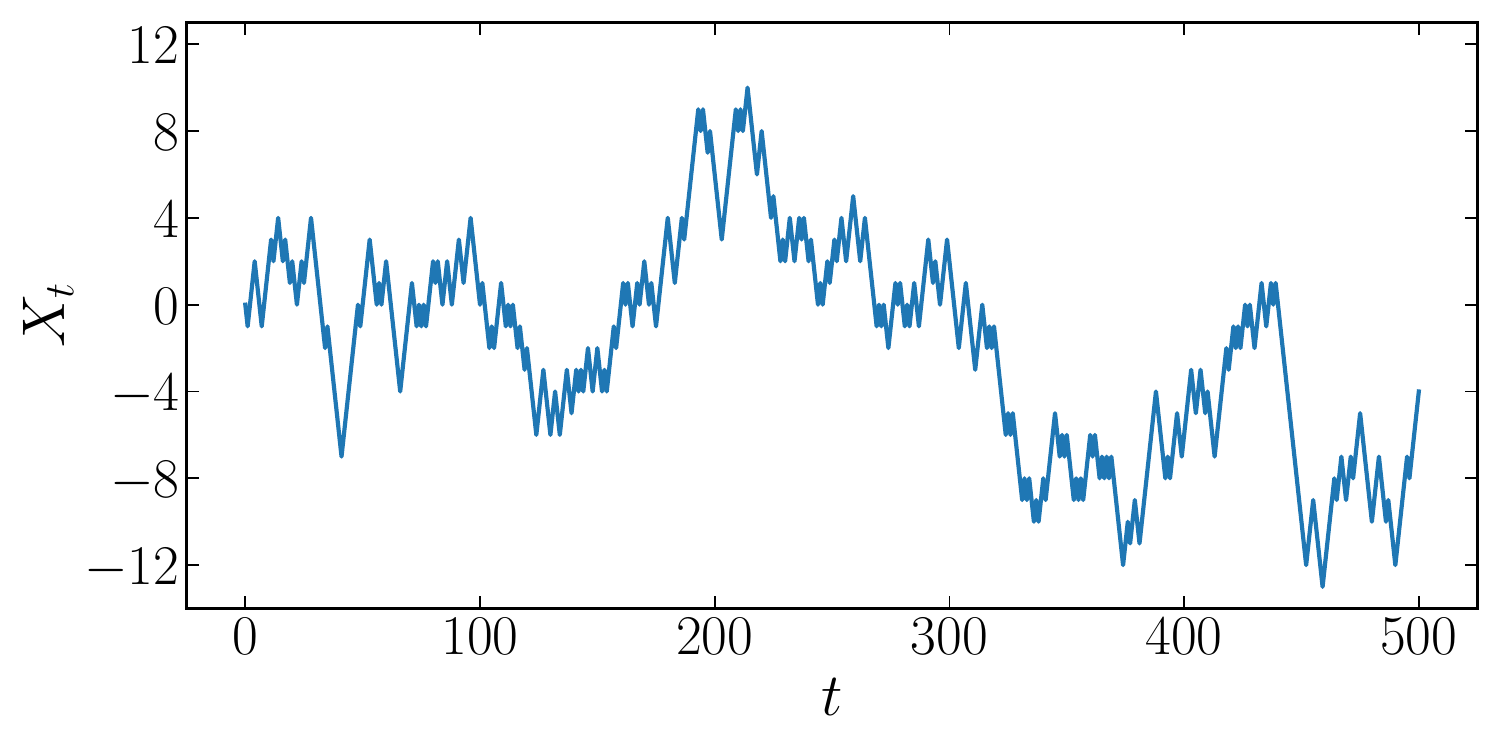}
\caption{\label{fig:1} Trajectory of an unbiased random walk on $\mathbb{Z}$ where $T=\{0,1,...,500\}$. }
\end{figure}

\noindent Based on this property, Markov processes are said to describe memoryless dynamics since the system state at time $t_n$ depends only on 
its immediate past, and not on all previous states: the system `forgets' its history and exhibits no memory. 
This type of process can physically occur when there is a large degree of separation between the system evolution timescale $\tau_R$ and correlation time $\tau_B$.
Consider, for example, a colloidal particle in a fluid undergoing Brownian motion. 
Assuming the particle's position takes discrete values, the particle dynamics can be approximated as an 
unbiased random walk\footnote{While Brownian motion is typically formulated as a continuous-time process, it can be discretized in time by sampling
the particle's position at fixed time intervals $\Delta t\sim \frac{t_n-t_1}{n}$ ($n\rightarrow \infty$ recovers the continuous limit).
This discretization leads to a process known as a random walk.}---an example of a discrete-time Markov chain---if 
(i) particle-fluid collisions are weak, (ii) the rate at which these
collisions occur $\tau^{-1}_B$ is much larger than the particle's relaxation rate $\tau^{-1}_R$ (as determined by the friction coefficient and particle's mass \cite{Pathria2021}),
and (iii) the time step $\Delta t$ over which the process is coarse-grained fulfils $\Delta t\gg \tau_B$.
An example trajectory of such a process is shown in Fig. \ref{fig:1}. 

To make our discussion more concrete, let us assume a dynamics where $X_t$ takes discrete integer values $\mathbb{N}_d = \{1,2,...,d\}$, and introduce ($t\geq s\geq 0$) \cite{Cantuck2024}
\begin{align}
    \mathbb{P}^d &= \bigg\{\boldsymbol{p}  :  \sum_{i=1}^d p_i = 1, \, p_i \geq 0 \bigg\}, \\
    \boldsymbol{p}(t) &= (P_1(X_t=1), P_1(X_t=2), ..., P_1(X_t=d))^T,   \\ 
    T_{ij}(t,s) &:= P_{1|1}(X_t=i|X_s=j), \label{eq:T}
\end{align}
where $T_{ij}(t,s)$ are the coefficients of a so-called transition or stochastic matrix. 
By construction, the transition matrix $\boldsymbol{T}(t,s)$ describes the time evolution of one-point probabilities via
\begin{equation}\label{eq:one_point}
\boldsymbol{p}(t) = \boldsymbol{T}(t,s)\boldsymbol{p}(s)
\end{equation}
and represents a positive map on $\mathbb{P}^d$ with the properties
\begin{equation}\label{eq:T_prop}
    T_{ij}(t,s)\geq 0, \quad T_{ij}(t,t) = \delta_{ij}, \quad \sum_iT_{ij}(t,s) = 1.
\end{equation}
We remark that a process is called time homogenous if the family of transition matrices $\boldsymbol{T}(t,s)$ depends only the difference in their time arguments,
\begin{equation}\label{eq:time_homo}
    \boldsymbol{T}(t+\tau,s+\tau) =  \boldsymbol{T}(t,s) = \boldsymbol{T}(t-s,0), \quad \forall \tau\in\mathbb{R}.
\end{equation}

Consider now the three-point probability distribution $P_3(x_{t}, x_{s}, x_{u})$ of a Markov chain for any $t\geq s\geq u$.
According to the Markov property \eqref{eq:Markov}, this distribution can be factorized as
\begin{align}
    P_3(x_{t}, x_{s}, x_{u}) &= P_{1|1}(x_{t}|x_{s},x_{u})P_1(x_{s}, x_{u}) \nonumber \\
                             &= T_{x_{t}x_{s}}(t,s)T_{x_{t}x_{s}}(s,u)P_1(x_{u}),
\end{align}
so that by marginalizing over $X_{s}$ and dividing through by $P_1(x_{u})$, we obtain the composition law
\begin{equation}\label{eq:CKE}
    \boldsymbol{T}(t,u) = \boldsymbol{T}(t,s)\boldsymbol{T}(s,u)
\end{equation}
known as the Chapman-Kolmogorov equation (CKE).
Importantly, this composition law establishes a necessary condition that the transition matrix \eqref{eq:T} must fulfil
to define a Markov process.
It is conversely true that all $n$-point probability distributions of a Markov process can be reconstructed from
an initial probability vector $\boldsymbol{p}(0)$ and transition matrix $\boldsymbol{T}(t,s)$ satisfying the CKE \cite{Breuer2016,Rivas2014}.

\begin{theorem}
A Markov process $\{X_t : t\in T\}$ is uniquely defined by an initial probability vector $\boldsymbol{p}(0)$ and transition matrix $\boldsymbol{T}(t,s)$ with the properties \eqref{eq:T_prop} and \eqref{eq:CKE}.
\end{theorem}

For $n>1$, define
\begin{equation}\label{eq:joint_prob}
    P_n(x_{t_n},...,x_{t_1}) = \prod^n_{j=2}T_{x_{t_j}x_{t_{j-1}}}(t_j,t_{j-1})p_{x_{t_1}}. 
\end{equation}
By construction, $P_{1|n}(x_{t_n}|x_{t_{n-1}},...,x_{t_1})$ obeys the Markov property \eqref{eq:Markov} since $T_{x_{t_j}x_{t_{j-1}}}(t_j,t_{j-1})$ represent two-point conditional probabilities. 
Furthermore, it can be shown using the CKE that the family of joint distributions \eqref{eq:joint_prob} satisfy the Kolmogorov consistency conditions (see Exercise 1).
It therefore follows that a given $\boldsymbol{p}(0)$ and $\boldsymbol{T}(t,s)$ uniquely characterize a Markov process. \qed 

\subsubsection{P-divisible processes}

We now introduce a more general class of processes based on the notion of P-divisible maps.
To this end, consider the one-parameter family of transition matrices $\{\boldsymbol{T}(t,0) : t\geq 0\}$ describing the time evolution of a probability vector $\boldsymbol{p}(t)$ according to
\begin{equation}\label{eq:stoch_map}
    \boldsymbol{p}(t) = \boldsymbol{T}(t,0)\boldsymbol{p}(0).
\end{equation}
Provided that the map $\boldsymbol{T}(t,0):\mathbb{P}^d\rightarrow\mathbb{P}^d$ is invertible, i.e., the matrix inverse $\boldsymbol{T}(t,0)^{-1}$ exists for all $t\geq 0$, it is possible to algebraically construct a two-parameter family of maps obeying the composition law
\begin{equation}\label{eq:int_map}
    \boldsymbol{T}(t,0) = \boldsymbol{V}(t,s)\boldsymbol{T}(s,0),
\end{equation}
with $\boldsymbol{V}(t,s) := \boldsymbol{T}(t,0)\boldsymbol{T}(s,0)^{-1}$. 
Such a construction leads to the following notion of a P-divisible process.

\begin{definition}[P-divisible process]
    A P-divisible process is a stochastic process $\{X_t : t\in T\}$ for which the intermediate maps \eqref{eq:int_map} 
    satisfy 
    \begin{equation}\label{eq:p_div}
        \sum_iV_{ij}(t,s) = 1, \quad V_{ij}(t,s)\geq 0,
    \end{equation}
    for all $t\geq s\geq 0$.
\end{definition}

\noindent An immediate consequence of the property \eqref{eq:p_div} is that any transition matrix $\boldsymbol{T}(t,s)$ obeying the CKE will also define a P-divisible process. 
In particular, by setting $u=0$ in \eqref{eq:CKE}, the CKE becomes equivalent to \eqref{eq:int_map} with $\boldsymbol{V}(t,s)=\boldsymbol{T}(t,s)$, implying that all Markov processes are P-divisible.
On the other hand, it should be stressed that while the matrix elements $V_{ij}(t,s)$ of a P-divisible map satisfy the same positivity and normalization as $T_{ij}(t,s)$, 
the condition \eqref{eq:p_div} does not guarantee that $V_{ij}(t,s)$ can be interpreted as a genuine conditional probabilities. 
This can be illustrated through the following example of a P-divisible process that is also non-Markovian (see also Definition 2.2 in \cite{Rivas2014}). 
\begin{example}
    Consider a two-state process with sample space $\Omega=\{0,1\}^3$ and $\boldsymbol{p}(0) = (\tfrac{2}{3}, \tfrac{1}{3})^T$, where $t>s>0$. 
    Let 
    \begin{equation}
        \boldsymbol{T}(s,0) = 
        \begin{pmatrix}
              3/4 & 1/4 \\
              1/4 & 3/4
        \end{pmatrix}
    \end{equation}
    define the transition matrix for the first time step $0\rightarrow s$. This yields the one-point probabilites 
    \begin{equation}
        \boldsymbol{p}(s) = 
        \begin{pmatrix}
            3/4 & 1/4 \\
            1/4 & 3/4
        \end{pmatrix} 
        \begin{pmatrix}
            2/3 \\
            1/3
        \end{pmatrix}
        =
        \begin{pmatrix}
            7/12 \\
            5/12
        \end{pmatrix}.
    \end{equation}
    Next, we introduce a non-Markovian transition rule in terms of the conditional probablities $P_{1|1}(X_t=0|X_0=0) = 3/4$, and $P_{1|1}(X_t=0|X_0=1) = \tfrac{1}{4}$, such that (see \eqref{eq:T})
    \begin{equation}
        \boldsymbol{T}(t,0) = 
        \begin{pmatrix}
            3/4 & 1/4 \\
            1/4 & 3/4
        \end{pmatrix}
        = \boldsymbol{T}(s,0).
    \end{equation}%
    This enables us to compute the intermediate map
    \begin{equation}
        \boldsymbol{V}(t,s) = \boldsymbol{T}(t,0)\boldsymbol{T}(s,0)^{-1} = 
        \begin{pmatrix}
            1 & 0 \\
            0 & 1      
        \end{pmatrix},
    \end{equation} 
    as well as the marginals
    \begin{equation}
        \boldsymbol{p}(t) = \boldsymbol{V}(t,s)\boldsymbol{p}(s) =
        \begin{pmatrix}
            7/12 \\    
            5/12
        \end{pmatrix}.
    \end{equation}
    Hence, the process is P-divisible. Now compare $ \boldsymbol{V}(t,s)$ to the transition matrix $\boldsymbol{T}(t,s)$, which can be evaluated via the probability rule $P_{1|1}(X_t=i|X_s=j) = \sum_kP_{1|1}(X_t=i|X_0=k)P_{1|1}(X_0=k|X_s=j)$,
    \begin{equation}\label{eq:T_ts}
        \boldsymbol{T}(t,s) = 
        \begin{pmatrix}
            19/28 & 9/20 \\
            9/28 & 11/20
        \end{pmatrix}
        \neq \boldsymbol{V}(t,s).
    \end{equation}
    This highlights how the intermediate map $\boldsymbol{V}(t,s)$ can differ from $\boldsymbol{T}(t,s)$. 
    Consequently, the matrix elements of $\boldsymbol{V}(t,s)$ cannot be interpreted as conditional probabilities despite reproducing the same one-point process as $\boldsymbol{T}(t,s)$. 
    We also remark that while the process is P-divisible, it violates the CKE:
    \begin{equation}\label{eq:CKE_violation}
        \boldsymbol{T}(t,s)\boldsymbol{T}(s,0) = 
        \begin{pmatrix}
            19/28 & 9/20 \\
            9/28 & 11/20
        \end{pmatrix}
        \begin{pmatrix}
            3/4 & 1/4 \\
            1/4 & 3/4
        \end{pmatrix} \\
        =
        \begin{pmatrix}
            \frac{87}{140} & \frac{71}{140} \\[1ex] 
            \frac{53}{140} & \frac{69}{140}
        \end{pmatrix}
        \neq \boldsymbol{T}(t,0).
    \end{equation} 
\end{example}

\bigskip

\noindent All in all, our analysis infers the existence of a hierarchy
\begin{equation}\label{eq:hier}
\textrm{Markov} \subset \textrm{CKE} \subset \textrm{P-divisible}.
\end{equation}
It should be stressed, however, that the ability to distinguish these types of processes generally relies on knowledge of the complete hierarchy of conditional probabilities \eqref{eq:st_cond_prob} (see e.g., \eqref{eq:T_ts} and \eqref{eq:CKE_violation}).  
To illustrate this, suppose we are only given access to the transition matrix $\boldsymbol{T}(t,0)$ used to evolve one-point probabilites $\boldsymbol{p}(t)$.  
While this allows us to assess whether the process is P-divisible, 
it tells us nothing about the conditional probabilities $\boldsymbol{T}(t,s)$, which can only be extracted from higher-order joint probability distributions. Thus, P-divisibility and CKE in such cases are indistinguishable \cite{Rivas2014}.

\begin{proposition}[P-divisibility = CKE]
    P-divisibility and CKE are operationally equivalent at the level of one-point probabilites $\boldsymbol{p}(t)$, in that they
    describe the same composition law $\boldsymbol{T}(t,u)=\boldsymbol{T}(t,s)\boldsymbol{T}(s,u)$.
\end{proposition}

Consider a process characterized by an initial distribution $\boldsymbol{p}(0)$ and family of maps $\boldsymbol{T}(t,0)$. 
If the process is P-divisible, then it is equally possible to propagate the one-point probabilites $\boldsymbol{p}(t)$ through a transition matrix $\boldsymbol{T}(t,s)\overset{!}{=}\boldsymbol{V}(t,s)$ satisfying the CKE \cite{Vacchini2011}. In particular, 
\begin{equation}
    \boldsymbol{T}(t,0) = \boldsymbol{V}(t,s) \boldsymbol{T}(s,0) = \underbrace{\boldsymbol{V}(t,s)\boldsymbol{V}(s,u)}_{=\boldsymbol{V}(t,u)}\boldsymbol{T}(u,0),
\end{equation}
such that 
\begin{equation}\label{eq:p_div_cke}
    \boldsymbol{T}(t,s) \overset{!}{=} \boldsymbol{V}(t,s) \,\, \Rightarrow \,\, \boldsymbol{T}(t,u) = \boldsymbol{T}(t,s)\boldsymbol{T}(s,u), 
\end{equation}
where any inconsistencies with the identification \eqref{eq:p_div_cke} only appear in higher-order joint probability distributions. \qed

\bigskip

\noindent By extension, it also possible to construct a Markov process whose $n$-point probabilites \eqref{eq:joint_prob} reproduce the same marginals $\boldsymbol{p}(t)$ obtained through \eqref{eq:p_div_cke}.
This implies that P-divisibility and Markovianity are operationally equivalent when restricted to one-point probabilities, since they impose the same conditions on the intermediate map $\boldsymbol{V}(t,s)$ \cite{Rivas2014}.

\begin{definition}[Markovianity = P-divisibility]\label{def:3}
A stochastic process solely characterized by $\{\boldsymbol{T}(t,0) : t\in T\}$ and $\boldsymbol{p}(0)$ is said to be Markovian if $\boldsymbol{T}(t,0)$ is P-divisible.
\end{definition}

\noindent Finally, an important quantity in the study of P-divisible processes is the Kolmogorov distance
\begin{equation}\label{eq:K_dist}
        D_{\rm K}(\boldsymbol{p}^1,\boldsymbol{p}^2) = \frac{1}{2}\|\boldsymbol{p}^1 - \boldsymbol{p}^2\|_1,
\end{equation}
where $\|\boldsymbol{p}\|_1 = \sum^d_{i=1}|p_i|$ denotes the $\ell_1$-norm\footnote{To simplify notation, we use $\|\cdot\|_1$ to denote both the $\ell_1$-norm and the trace norm, where the intended meaning should be clear from the context.} of $\boldsymbol{p}$, and $\boldsymbol{p}^{1,2}\in\mathbb{P}^d$.
The Kolmogorov distance can be shown to be zero if and only if $\boldsymbol{p}^1=\boldsymbol{p}^2$, and that its maximum value ${\rm max}_{\boldsymbol{p}^{1,2}}D_{\rm K}(\boldsymbol{p}^1,\boldsymbol{p}^2) = 1$ is attained for orthogonal probability vectors.   
It also the case that 
\begin{equation}\label{eq:kd_contract}
D_{\rm K}(\boldsymbol{p}^1(t), \boldsymbol{p}^2(t))\leq D_{\rm K}(\boldsymbol{p}^1(0),\boldsymbol{p}^2(0))
\end{equation} 
for all $t\geq 0$ through the property of $\boldsymbol{T}(t,0)$ being contractive, i.e., $\|\boldsymbol{T}\boldsymbol{x}\||_1\leq\|\boldsymbol{x}\|_1$, with $\boldsymbol{x}\in\mathbb{R}^d$ \cite{Chruscinski2022}.  
Furthermore, the Kolmogorov distance can be shown to be a strictly decreasing quantity in time provided that the underlying process is P-divisible \cite{Vacchini2011}.  

\begin{proposition}
    Let $\boldsymbol{T}(t,0)$ be the transition matrix of a P-divisible stochastic process with $\boldsymbol{p}(t)=\boldsymbol{T}(t,0)\boldsymbol{p}(0)$. 
    Then the Kolmogorov distance \eqref{eq:K_dist} satisfies the inequality 
    \begin{equation}\label{eq:mdkd}
        D_{\rm K}(\boldsymbol{p}^1(t), \boldsymbol{p}^2(t)) \leq D_{\rm K}(\boldsymbol{p}^1(s), \boldsymbol{p}^2(s)),
    \end{equation}
    for all $t\geq s \geq 0$. 
\end{proposition}

We apply the definitions of the Kolmogorov distance \eqref{eq:K_dist} and intermediate map \eqref{eq:p_div}:
\begin{align}
    \frac{1}{2}\sum_i|p^1_i(t) - p^2_i(t)| &= \frac{1}{2}\sum_i\Big|\sum_jV_{ij}(t,s)[p^1_j(s) - p^2_j(s)]\Big|  \nonumber\\
                                           &\leq \frac{1}{2}\sum_{i,j}V_{ij}(t,s)\big|p^1_j(s) - p^2_j(s)\big| \nonumber\\
                                           &= \frac{1}{2}\sum_j|p^1_j(s)-p^2_j(s)|.
\end{align}
The second line follows from the triangle inequality and positivity of $V_{ij}(t,s)$, while the third line uses the normalization $\sum_iV_{ij}(t,s)=1$. 
Thus from $\tfrac{1}{2}\sum_j|p^1_j(s)-p^2_j(s)| = D_{\rm K}(\boldsymbol{p}^1(s), \boldsymbol{p}^2(s))$, we obtain the stated result.  \qed 

\bigskip

\noindent According to Definition \ref{def:3}, a process may therefore be considered Markovian if the Kolmogorov distance monotonically decreases under the map $\boldsymbol{T}(t,0)$.
We further remark that the violation of \eqref{eq:mdkd} does not contradict the contractivity property \eqref{eq:kd_contract}: 
while the Kolmogorov distance may temporarily increase above the value $D_{\rm K}(\boldsymbol{p}^1(s), \boldsymbol{p}^2(s))$ at some time $t>s$, 
it will always be bounded from above by its initial value (i.e., its value cannot grow starting at time $t=0^+$).

\subsubsection{Markov semigroups and differential CKE}

For time homogenous Markov chains, the CKE reproduces the following semigroup relation
\begin{equation}\label{eq:cke_semigroup}
\boldsymbol{T}(t+s) = \boldsymbol{T}(t)\boldsymbol{T}(s),
\end{equation}
where $\boldsymbol{T}(t):=\boldsymbol{T}(t,0)$ and $\boldsymbol{T}(0) = \mathbb{I}$. 
Based on the connection between the CKE and Markov processes, any family of maps $\{\boldsymbol{T}(t):t\in T\}$ obeying 
\eqref{eq:cke_semigroup} is said to define a Markov semigroup. 
Let us for convenience sake now assume such processes may be extended to the time continuous case $t\in\mathbb{R}^+_0$.
Assuming the map $\boldsymbol{T}(t)$ to be continuous at $t=0$, 
one can then apply \eqref{eq:cke_semigroup} to define the so-called Kolmogorov generator 
$\boldsymbol{L} = \lim_{\epsilon\rightarrow0^+}\frac{\boldsymbol{T}(\epsilon)-\mathbb{I}}{\epsilon}$, such that 
\begin{equation}
    \frac{d}{dt}\boldsymbol{T}(t) = \boldsymbol{L}\boldsymbol{T}(t) \,\, \Rightarrow \,\, \boldsymbol{T}(t) = e^{\boldsymbol{L}t},
\end{equation}
where the exponential of $\boldsymbol{L}$ is well-defined due to $\boldsymbol{T}(t)$ being bounded. 
This leads to the following equation of motion for the probability vector \eqref{eq:one_point}, 
\begin{equation}\label{eq:diff_cke}
    \frac{d}{dt}\boldsymbol{p}(t) = \boldsymbol{L}\boldsymbol{p}(t),
\end{equation}
also known as the Markovian master equation or differential CKE \cite{Vacchini2024}. 
To maintain the positivity and normalization of $\boldsymbol{p}(t)$, the coefficients of the Kolmogorov generator $L_{ij}$ must satisfy \cite{Kampen2007}
\begin{equation}
    L_{ij} \geq 0, \qquad \sum^d_{i=1}L_{ij} = 0.
\end{equation} 
These coefficients may similarly be expressed as $L_{ij} = W_{ij} - W_j\delta_{ij}$, where $W_{ij}$ ($W_{ji}$) represent the forward (backward) transition rates for the process $j\rightarrow i$ ($i\rightarrow j$),
and $W_j = \sum_kW_{jk}$ \cite{Chruscinski2022}. 
Accordingly, the master equation \eqref{eq:diff_cke} can be written component-wise as \cite{Schaller2014}
\begin{equation}
    \frac{d}{dt}p_i(t) = \sum_j\Big[W_{ij}p_j(t) - W_{ji}p_i(t)\Big]. 
\end{equation} 
It is worth noting that this form of differential CKE bears a close relation to the Gorini-Kossakowski-Sudarshan-Lindblad master equation, which will later be discussed in the context of quantum Markov processes.

\section{Open quantum systems}\label{sec:3}

In this section we review some aspects of open quantum systems theory, 
focusing on the formulation of completely positive and trace-preserving (CPTP) maps and the notion of divisibility of such maps.  
Since much of the introductory material on this topic is covered in \cite{Merkli2026}, familiarity with concepts such as density matrices, composite Hilbert spaces, e.t.c., is assumed. 
A more extensive review of the theory may also be found in \cite{BP2002, Rivas2010,Chruscinski2022,Vacchini2024}.

We start with the conventional formulation of quantum states on a finite dimensional complex vector space $\mathcal{H}$. 
The set of all possible states of a quantum system is represented by the convex set of trace-class operators\footnote{A trace-class operator $O$ is defined such that $\|O\|_1<\infty$, where $\|O\|_1:= {\rm Tr}|O| = \sum_n\langle n|\sqrt{O^{\dagger}O}|n\rangle$ denotes the trace norm, and $\{|n\rangle\}$ is an orthonormal basis for $\mathcal{H}$.} on $\mathcal{H}$ satisfying the properties
\begin{equation}\label{eq:set_dm}
    \mathcal{S}(\mathcal{H}) = \{\rho\,|\,\rho = \rho^{\dagger}, \,\rho\succeq 0, \, {\rm Tr}[\rho]=1\},
\end{equation}
where the operators $\rho$ are known as density matrices. 
The Hermiticity of $\rho$ is required to ensure that such density matrices are diagonalizable, 
while the non-negativity and unit trace properties ensure the eigenvalues of $\rho$ may be interpreted as probabilities. 
An immediate consequence of this definition is that the operators $\rho$ obey the inequality 
\begin{equation}
    {\rm Tr}[\rho^2] \leq {\rm Tr}[\rho] = 1.
\end{equation}
Density operators which saturate this inequality are known as pure states, and are always of the form of rank-one matrices $\rho = |\psi\rangle\langle\psi|$,
where $\psi\in\mathcal{H}$. 
Alternatively, mixed states $\rho = \sum_iw_i|\psi_i\rangle\langle\psi_i|$ are obtained as convex mixtures of pure states $|\psi_i\rangle\langle\psi_i|$ 
with non-negative weights satisfying $\sum_iw_i =1$.
\begin{figure}[t!]
\centering
\includegraphics[scale=0.36]{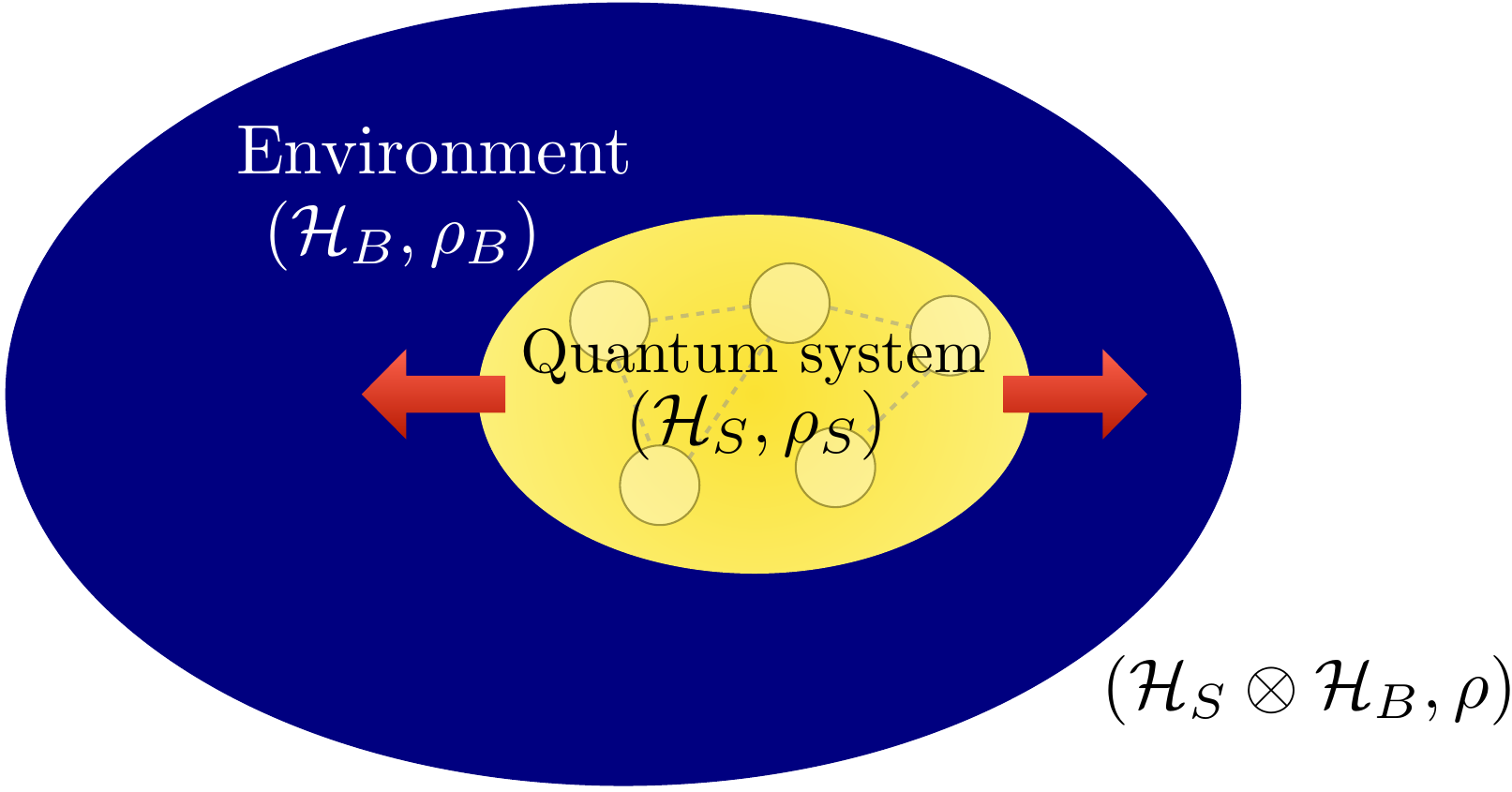}
\caption{General framework for open quantum systems. 
The composite space $\mathcal{H}$ on which all physical states of $S+B$ are defined has the tensor product structure $\mathcal{H}=\mathcal{H}_S\otimes\mathcal{H}_B$. 
If $\rho$ belongs to the set of physical states $\mathcal{S}(\mathcal{H})$, then the marginal states of $S$ and $B$ can be obtained via the partial trace. 
The reduced density matrices $\rho_S$ and $\rho_B$ fully characterize the one-point statistics of the respective subsystems.}
\label{fig:2}
\end{figure}

The composite Hilbert space of an open system $S$ and environment (bath) $B$ has the tensor product structure 
\begin{equation}
    \mathcal{H} = \mathcal{H}_S\otimes \mathcal{H}_B.
\end{equation}
Accordingly, the reduced density matrix $\rho_S$ representing the marginal or reduced state of $S$ can be obtained as
\begin{equation}\label{eq:rdm}
   \rho_S := {\rm Tr}_B[\rho],
\end{equation}
where ${\rm Tr}_B:\mathcal{B}(\mathcal{H})\rightarrow \mathcal{B}(\mathcal{H}_S)$ denotes the partial trace, and $\mathcal{S}(\mathcal{H})\subset \mathcal{B}(\mathcal{H})$.
This operation defines a linear map from the space of bounded operators $\mathcal{B}(\mathcal{H})$ on the open system and environment to the subspace of bounded operators acting on $S$ alone
(i.e., it maps between spaces of different dimension). 
Since the partial trace further represents a CPTP map \cite{Lidar2020}, the reduced density matrix \eqref{eq:rdm} 
can be treated as a valid quantum state in same vein as $\rho$
insofar that it contains the same statistical information as $\rho$ when limited to performing local measurements on the subsystem $S$ only\footnote{In other words, the partial trace ${\rm Tr}_B$ may also be regarded as a linear map from $\mathcal{S}(\mathcal{H})$ to $\mathcal{S}(\mathcal{H_S})$.} \cite{Rivas2010}.

\subsection{Time evolution and CPTP maps}

Assuming the composite system $S+B$ to be isolated (see Fig. \ref{fig:2}), the time evolution of the state $\rho(0)=\rho$ is prescribed through ($\hbar=1$ units)
\begin{equation}
    \rho(t) = U(t)\rho(0)U^{\dagger}(t).
\end{equation}
Here, $U(t) = e^{-iHt}$ is the unitary time evolution operator, and 
\begin{equation}\label{eq:H}
   H = H_S + H_B + H_I
\end{equation}
is the total Hamiltonian of the open quantum system and environment. 
In particular, $H_S$ and $H_B$ represent the free Hamiltonians of $S$ and $B$ when they are uncoupled, and $H_I$ is an interaction term.
Suppose that the initial state $\rho(0)$ is factorized as
\begin{equation}\label{eq:initial}
    \rho(0)=\rho_S(0)\otimes\rho_B,
\end{equation}
which occurs when the open system and environment are initially uncorrelated (or statistically independent). 
In this case, the marginal state of $S$ after time $t\geq 0$ reads 
\begin{equation}\label{eq:rsd}
    \rho_S(t) = {\rm Tr}_B[U(t)(\rho_S(0)\otimes\rho_B)U^{\dagger}(t)].
\end{equation}
Using of the spectral decomposition $\rho_B=\sum_k\lambda_k|k\rangle\langle k|_B$, where $\{|k\rangle_B\}$ defines an orthonormal basis for $\mathcal{H}_B$, we obtain 
\begin{align}\label{eq:kraus}
    \rho_S(t) &= \Phi_t\rho_S(0) \nonumber \\
              &:= \sum_{\alpha}K_{\alpha}(t)\rho_S(0)K^{\dagger}_{\alpha}(t).
\end{align}
Here, $K_{\alpha=(j,k)}(t) = \sqrt{\lambda_k}\langle j|U(t)|k\rangle_B$ are system operators $K_{\alpha}: \mathcal{H}_S\rightarrow\mathcal{H}_S$, 
also known as Kraus operators \cite{Kraus1983}.
By definition, these operators satisfy the completeness relation 
\begin{equation}\label{eq:complete}
       \sum_{\alpha}K^{\dagger}_{\alpha}(t)K_{\alpha}(t) = \mathbb{I}_S.
\end{equation}
One may regard \eqref{eq:kraus} as defining a one-parameter family of linear maps from the space of bounded system operators (including density matrices) onto itself:
\begin{equation}\label{eq:map}
\Phi_t:\mathcal{B}(\mathcal{H}_S)\rightarrow\mathcal{B}(\mathcal{H}_S).
\end{equation}
Remarkably, the map $\Phi_t$ possesses a mathematical structure that is both necessary and sufficient for it to describe any physical operation satisfying three criteria: 
(i) trace preservation, (ii) Hermiticity, and (iii) complete positivity \cite{Lidar2020}.
To illustrate this, we can first prove the weaker result that maps with a Kraus structure \eqref{eq:kraus} are trace preserving, Hermitian, and (iv) positive.

\begin{lemma}[Positive trace preserving maps]\label{lemma:1}
    Let $\Phi$ define a map on $\mathcal{B}(\mathcal{H}_S)$ with $d={\rm dim}\,\mathcal{H}_S$, and $X\in \mathcal{B}(\mathcal{H_S})$. 
    Then the Kraus representation $\Phi(X) = \sum^{d^2}_{\alpha=1}K_{\alpha}XK^{\dagger}_{\alpha}$ is sufficient for $\Phi$ to satisfy (i), (ii), and (iv). 
\end{lemma}
These first two properties can be easily verified as a result of 
the completeness of the Kraus operators \eqref{eq:complete} and the sesquilinear structure of $\Phi(X)$.
\begin{itemize}
    \item[(i)] Trace preservation:
    \begin{equation*}
    {\rm Tr}[\Phi(X)] = \sum_{\alpha}{\rm Tr}[K_{\alpha}XK^{\dagger}_{\alpha}] = {\rm Tr}\bigg[\bigg(\sum_{\alpha}K^{\dagger}_{\alpha}K_{\alpha}\bigg)X\bigg] = {\rm Tr}[X],
    \end{equation*}
    \item[(ii)] Hermiticity:
    \begin{equation*}
    \Phi(X^{\dagger}) = \sum_{\alpha}K_{\alpha}X^{\dagger}K^{\dagger}_{\alpha} = \sum_{\alpha}(K_{\alpha}XK^{\dagger}_{\alpha})^{\dagger} = \Phi(X)^{\dagger}.
    \end{equation*}
\end{itemize}
The positivity of $\Phi$ requires that $\Phi(X)$ is a positive operator as long as $X$ is positive, namely $\langle \psi|\Phi(X)|\psi\rangle \geq 0$, $\forall \psi\in\mathcal{H}_S$. 
\begin{itemize}
    \item[(iv)] Positivity:
\begin{equation*}
    \langle \psi|\Phi(X)|\psi\rangle =  \sum_{\alpha}\langle \psi|K_{\alpha} X K^{\dagger}_{\alpha}|\psi\rangle = \sum_{\alpha}\langle \phi_{\alpha}|X|\phi_{\alpha}\rangle \geq 0. 
\end{equation*}
\end{itemize}
On the other hand, complete positivity requires that the dilated map $\Phi\otimes\mathcal{I}^{(n)}_A$ acting on both $S$ and an ancillary system $A$ 
is positive $\forall n\geq 1$, where $n={\rm dim}\,\mathcal{H}_A$, 
and $\mathcal{I}_A$ is the identity map, $\mathcal{I}_A(X)=X$.
In particular, we call a map $\Phi$ $n$-positive if its dilation $\Phi\otimes\mathcal{I}^{(n)}_A$ is a positive map.
Since positivity coincides with complete positivity when $n=1$, it follows that a map which is CP is also positive, 
but not vice versa\footnote{More intuitively, the difference between complete positivity and positivity is that the former prevents the marginal states of bipartite open systems being nonpositive. For example, if $S=S_1+S_2$, the marginal state ${\rm Tr}_{S_2}[\Phi(\rho_{S_1S_2})]$ could have negative eigenvalues if $\Phi$ were only positive, meaning it could not represent a valid quantum state.}. 

Importantly, the complete positivity of $\Phi_t$ is guaranteed through the following theorem of Kraus \cite{Kraus1983}. 
\begin{theorem}[Kraus theorem on CPTP maps]
    Let $\Phi$ define a CPTP map on $\mathcal{B}(\mathcal{H}_S)$ with $d={\rm dim}\,\mathcal{H}_S$.
    Then there exists a family of Kraus operators $\{K_{\alpha}\}^{d^2}_{\alpha=1}$ where $K_{\alpha}\in\mathcal{B}(\mathcal{H}_S)$ and $\sum_{\alpha}K^{\dagger}_{\alpha}K_{\alpha}=\mathbb{I}_S$, such that $\Phi$ can be expressed in the form 
    \begin{equation}\label{eq:CPTP}
        \Phi(X) = \sum_{\alpha}K_{\alpha}XK^{\dagger}_{\alpha}.
    \end{equation}
    Conversely, given a family of Kraus operators $\{K_{\alpha}\}^{d^2}_{\alpha=1}$ where $\sum_{\alpha}K^{\dagger}_{\alpha}K_{\alpha}=\mathbb{I}_S$, then the map \eqref{eq:CPTP} is CPTP.
\end{theorem}

A comprehensive proof can be found in \cite{Merkli2026}. \qed

\medskip

\noindent Accordingly, maps exhibiting a Kraus structure \eqref{eq:kraus} are referred to as CPTP maps, since they satisfy (i)-(iii) by construction (Kraus $\Rightarrow$ CPTP).
The Kraus theorem also proves the reverse statement that all CPTP maps $\Phi$ admit a representation \eqref{eq:CPTP}, which can be specified in terms of any unitarily equivalent set of Kraus operators (CPTP $\Rightarrow$ Kraus).
In particular, let $K_{\alpha}=\sum_{\beta}u_{\alpha\beta}\tilde{K}_{\beta}$, where $u_{\alpha\beta}$ are the coefficients of a $d^2\times d^2$ unitary matrix.
Then the map $\Phi$ may equivalently be expressed as $\Phi(X) = \sum_{\alpha}\tilde{K}_{\alpha}X\tilde{K}^{\dagger}_{\alpha}$,
implying that the representation \eqref{eq:CPTP} is non-unique \cite{BP2002}.

The complete positivity of a linear map is also connected to its effect on maximally entangled states through a well-known theorem of Choi \cite{Choi1972}. 

\begin{theorem}[Choi theorem on CP maps]\label{theorem:3}
Let $\Phi$ define a map on $\mathcal{B}(\mathcal{H}_S)$ with $d={\rm dim}\,\mathcal{H}_S$. Then $\Phi$ is CP if and only if
\begin{equation}\label{eq:choi}
    C_{\Phi} = (\Phi \otimes \mathcal{I}^{(d)}_A)|\Phi^+_d\rangle\langle \Phi^+_d| \geq 0,
\end{equation}
where $C_{\Phi}$ is known as the Choi matrix of $\Phi$, and $|\Phi^+_d\rangle = \frac{1}{\sqrt{d}}\sum^{d}_{i=1}|i\rangle|i\rangle$ is a maximally entangled state in the composite space $\mathcal{H}_S\otimes\mathcal{H}_S$.
\end{theorem}

A proof can be found in e.g., \cite{Vacchini2024}. \qed

\bigskip

\noindent A key corollary of this result is that the complete positivity of a given map $\Phi$ can be verified through checking the eigenvalues of a suitably constructed matrix $C_{\Phi}$. 
This is explicitly demonstrated with the following example. 

\begin{example}\label{example:2}
    An example of a map that is positive but not completely positive is the reduction map \cite{Chruscinski2022}
     \begin{equation}\label{eq:red_map}
        \Phi(X) = {\rm Tr}(X)\mathbb{I}_S - X,
    \end{equation}
    which for a qubit system $\mathcal{H}_S={\rm span}\{|0\rangle,|1\rangle\}$, maps every pure state to its orthogonal complement.
    In the Bloch sphere representation, this is equivalent to reversing the sign of the Bloch vector $\boldsymbol{r}$ of the state 
    $\rho=\frac{1}{2}(\mathbb{I}_S+\boldsymbol{r}\cdot\boldsymbol{\sigma})$, i.e., $\Phi: \boldsymbol{r}\mapsto -\boldsymbol{r}$
    (states on antipodal points of the Bloch sphere are orthogonal).
    It is straightforward to verify that the map \eqref{eq:red_map} satisfies properties (i) and (iv):
    \begin{itemize}
        \item[(i)] $\Phi(X^{\dagger}) = {\rm Tr}(X)^*\mathbb{I}_S - X^{\dagger} = \Phi(X)^{\dagger}$.
        \item[(iv)] If $X\geq 0$, this guarantees $X$ to also be Hermitian, and so $X = \sum^d_{i=1}\lambda_i|i\rangle\langle i|$.
        Subsituting this decomposition into \eqref{eq:red_map} yields   
        \[
            \Phi(X) = \mathbb{I}_S - \sum_i\lambda_i|i\rangle\langle i| = \sum_i\bigg(\sum_{j\neq i}\lambda_j\bigg)|i\rangle\langle i|,
        \]
        where the eigenvalues $\sum_{i\neq j}\lambda_j$ are nonnegative since $\lambda_i\geq 0$, giving $\langle \psi|\Phi(X)|\psi\rangle\geq0$.
    \end{itemize}
    The complete positivity of $\Phi$  
    requires the Choi matrix \eqref{eq:choi} to be positive:
    \begin{equation*}
        C_{\Phi} = \frac{1}{d}\sum^d_{i,j=1}\Phi(|i\rangle\langle j|)\otimes |i\rangle\langle j| = \frac{1}{d}\sum_{i,j}\big[\delta_{ij}\mathbb{I}_S-|i\rangle\langle j|\big]\otimes |i\rangle\langle j|.
    \end{equation*}
    To prove $\Phi$ is non-CP, it suffices to show that $C_{\Phi}$ is nonpositive for $d=2$, using that the set of $d$-positive maps is entirely contained within $\{\Phi: (\Phi\otimes\mathcal{I}^{(2)}_A)X\geq0, \, \forall X\in\mathcal{B}(\mathcal{H}_S)\otimes\mathbb{C}^2\}$ \cite{Chruscinski2022}
    (i.e., if $\Phi\otimes\mathcal{I}^{(2)}_A$ is nonpositive, $\Phi\otimes\mathcal{I}^{(d)}_A$ is also guaranteed nonpositive).
    In this case we can write $C_{\Phi}$ in the ordered basis $\{|00\rangle,|01\rangle,|10\rangle,|11\rangle\}$ as
    \begin{equation*}
        C_{\Phi} = \frac{1}{2}\sum^1_{i,j=0}\big[\delta_{ij}\mathbb{I}_S-|i\rangle\langle j|\big]\otimes |i\rangle\langle j|
        =
        \frac{1}{2}
        \begin{pmatrix}
        0 & 0 & 0 & -1 \\
        0 & 1 & 0 & 0 \\
        0 & 0 & 1 & 0 \\
        -1 & 0 & 0 & 0
        \end{pmatrix},
    \end{equation*}
    whose spectrum is found to be $\{1/2,1/2,1/2,-1/2\}$.
    Hence, the negative eigenvalue $-1/2$ certifies that $\Phi$ is non-CP. 
\end{example}

\subsubsection{Trace distance under CPTP maps}\label{sec:trace_dist}

The trace distance between two quantum states $\rho^1,\rho^2\in\mathcal{S}(\mathcal{H}_S)$ is defined as \cite{Nielsen2012}
\begin{equation}\label{eq:td}
    D(\rho^1,\rho^2) = \frac{1}{2}\|\rho^1-\rho^2\|_1 = \frac{1}{2}\sum^d_{i=1}|\lambda_i|,
\end{equation}
where $\lambda_i$ denote the eigenvalues of the trace-class operator $\rho^1-\rho^2$, and $d={\rm dim}\,\mathcal{H}_S$. 
Notably, the trace distance defines a metric on the state space $\mathcal{S}(\mathcal{H}_S)$ in that it satisfies the properties:
\begin{itemize}
    \item[(i)] Permutation symmetry: $D(\rho^1,\rho^2) = D(\rho^2,\rho^1)$.
    \item[(ii)] Triangle inequality: $D(\rho^1,\rho^2) \leq D(\rho^1,\sigma) + D(\sigma,\rho^2)$, $\forall\sigma\in\mathcal{S}(\mathcal{H}_S)$.
    \item[(iii)] Distance property: $D(\rho^1,\rho^2) = 0$ $\Rightarrow$ $\rho^1=\rho^2$. 
\end{itemize}

The value of $D(\rho^1,\rho^2)$ is also maximized for any pair of states with orthogonal support: 
namely, $D(\rho^1,\rho^2) = 1$ if and only if $\rho^{1,2}=\sum_i\lambda^{1,2}_i|\lambda^{1,2}_i\rangle\langle \lambda^{1,2}_i|$, 
where $\{|\lambda^{1,2}_i\rangle\}_i$ span orthogonal subspaces of $\mathcal{H}_S$.
This property suggests the trace distance to be the quantum analogue of the Kolmogorov distance \eqref{eq:K_dist}, which vanishes for any pair of classical probability distributions with disjoint support.

Another important property of trace distance is that it is contractive under CPTP maps \cite{Ruskai1994}:
\begin{equation}\label{eq:td_cptp}
    D(\Phi(\rho^1),\Phi(\rho^2)) \leq D(\rho^1,\rho^2).
\end{equation}
In particular, this inequality is saturated if and only if $\Phi$ is unitary map, i.e., $\Phi\rho = U\rho U^{\dagger}$. 
An information theoretic interpretation of \eqref{eq:td_cptp} may also be considered in context of one-shot state discrimination as follows \cite{Rivas2014}. 
Suppose we are randomly given one of two possible states $\rho^{1,2}$ with probability $1/2$, and a positive operator-valued measure (POVM) $\{\Pi_k\}_{k\in\Omega}$ 
describing a generalized measurement (see \cite{Nielsen2012} for further details on POVMs).
We then perform a measurement and decide which state we have received based on the outcome $k$.
If $k$ belongs to the subset $\Omega_1\subset \Omega$, the state is taken as $\rho^1$, whereas if $k$ belongs to the complementary subset $\Omega_2 = \Omega\setminus\Omega_1$, the state is taken as $\rho^2$.
By choosing an optimal POVM, the maximum success probability of distinguishing $\rho^1$ and $\rho^2$ can be shown to equal $P_{\rm max}=\frac{1}{2}(1+D(\rho^1,\rho^2))$ \cite{Breuer2016,Breuer2009,Nielsen2012},
such that the trace distance can be interpreted as a measure of information regarding which state we received---or
that larger values of $D(\rho^1,\rho^2)$ indicates a higher distinguishability between the two states\footnote{To further illustrate this, consider the cases of parallel and orthogonal states where $D(\rho^1,\rho^2)=0$ and $D(\rho^1,\rho^2)=1$, respectively. 
For parallel states, all choices of POVM produce identical measurement statistics, and so we can only guess randomly whether we were given $\rho^1$ or $\rho^2$ with maximum success probability $P_{\rm max}=1/2$ (i.e., measurements reveal no information on the true state label 1 or 2).  
Conversely, orthogonal states are fully distinguishable, since by measuring the projection onto the support of $\rho^1$ and $\rho^2$, then outcome $k\in \Omega_1$ ($k\in\Omega_2$) infers that we received state $\rho^1$ ($\rho^2$) with probability $P_{\rm max}=1$. }. 
In turn, \eqref{eq:td_cptp} implies that the ability to distinguish $\rho^1$ and $\rho^2$ cannot increase under the action of a CPTP map.

\subsubsection{CP-divisible and P-divisible maps}\label{sec:3.1.2}

Recall that the notion of P-divisibility was introduced based on the positivity of the intermediate map $\boldsymbol{V}(t,s)$. 
Since the dynamical map $\Phi_t$ plays the same role as the stochastic matrix $\boldsymbol{T}(t,0)$ for quantum states, we may therefore introduce an analogous notion of CP-divisibility for CPTP maps. 
In particular, suppose that $\Phi_t$ is invertible for all $t\geq 0$.
We can then define a two-parameter family of maps $\Lambda_{t,s}$ obeying the composition law 
\begin{equation}\label{eq:CP_div}
    \Phi_t = \Lambda_{t,s}\Phi_s
\end{equation}
where $\Lambda_{t,s}:= \Phi_t\Phi_s^{-1}$, leading to the following definition.

\begin{definition}[CP-divisibility]\label{def:4}
    A CP-divisible map $\Phi_t$ is a dynamical map \eqref{eq:CP_div} where $\Lambda_{t,s}$ is CPTP for all $t\geq s\geq 0$. 
\end{definition}

\medskip

 \noindent Notice since $\Lambda_{t,s}$ is defined as the composition of two trace preserving maps, it is always trace preserving regardless of whether it is CP.  
The above definition of CP-divisibility also leads to the weaker notion of a P-divisible quantum map through Lemma \ref{lemma:1}.

\begin{definition}[P-divisibility]\label{def:5}
    A P-divisible map $\Phi_t$ is a dynamical map \eqref{eq:CP_div} where $\Lambda_{t,s}$ is positive and trace-preserving (PTP) for all $t\geq s\geq 0$. 
\end{definition}

\medskip 

\noindent Naturally, a map $\Phi_t$ which is CP-divisible is also P-divisible, given that all CP maps are positive by definition. 
We further remark that if $\{|i\rangle\}^d_{i=1}$ defines an orthonormal basis for $\mathcal{H}_S$, then one can define from $\Lambda_{t,s}$ the intermediate map
\begin{equation}\label{eq:qc_tm}
    V_{ij}(t,s) = \langle i|\Lambda_{t,s}(|j\rangle\langle j|)|i\rangle.
\end{equation}
In particular, if $\Lambda_{t,s}$ is CPTP this guarantees \eqref{eq:qc_tm} to fulfil the properties of a transition matrix:
\begin{align*}
    &\sum_iV_{ij}(t,s) = \sum_i\langle i|\Lambda_{t,s}(|j\rangle\langle j|)|i\rangle = {\rm Tr}[\Lambda_{t,s}(|j\rangle\langle j|)] = 1, \\
    &\langle i|\Lambda_{t,s}(|j\rangle\langle j|)|i\rangle \geq 0, \quad \forall i,j.
\end{align*}
This establishes that a P-divisible quantum dynamical map effectively induces an infinite number of P-divisible classical maps \cite{Benatti2024}, with the corresponding classical probability distributions in each basis set given by $p_i(t) = \langle i|\rho_S(t)|i\rangle$. 

The notion of divisibility may also considered in the context of quantum master equations of the form $\tfrac{d}{dt}\rho_S(t) = \mathcal{L}_t\rho_S(t)$, 
where $\mathcal{L}_t$ is known as the time-local generator of $\Phi_t$ \cite{BP2002,deVega2017}. 
In particular, suppose we have an invertible map $\Phi_t$ which may not necessarily be CPTP. Then $\mathcal{L}_t$ can be uniquely determined from (using $\frac{d}{dt}\rho_S(t)=\dot{\Phi}_t\rho_S(0)$ and $\Phi^{-1}_t\Phi_t=\mathcal{I}_S$)
\begin{equation}\label{eq:tlg}
    \frac{d}{dt}\rho_S(t) = (\dot{\Phi}_t\Phi^{-1}_t)\rho_S(t) := \mathcal{L}_t\rho_S(t),
\end{equation}
where $\mathcal{L}_t$ may also be defined from the intermediate map $\mathcal{L}_t = \lim_{\epsilon\rightarrow0^+}\tfrac{\Lambda_{t+\epsilon,t}-\mathcal{I}_s}{\epsilon}$.
Now, if the same map $\Phi_t$ is Hermitian and trace preserving for all $t\geq0$, it can be shown that the generator $\mathcal{L}_t$ must admit the canonical representation \cite{Vacchini2024,Hall2014}
\begin{equation}\label{eq:canonical}
    \mathcal{L}_t\rho = -i[H_0(t),\rho] + \sum^{d^2-1}_{j=1}\gamma_j(t)\Big[L_j(t)\rho L^{\dagger}_j(t) - \tfrac{1}{2}\{L^{\dagger}_j(t)L_j(t),\rho\}\Big].
\end{equation} 
Here, $H_0(t)$ is a Hermitian operator, and $\{L_j(t)\}^{d^2-1}_{j=1}$ are traceless operators which together with $L_0(t)=\tfrac{1}{\sqrt{d}}\mathbb{I}_S$, form a complete orthonormal basis of $\mathcal{B}(\mathcal{H}_S)$: 
\begin{equation}
    {\rm Tr}[L_j(t)] = 0, \qquad {\rm Tr}[L^{\dagger}_i(t)L_j(t)] = \delta_{ij}, \qquad i,j\in\{0,...,d^2-1\}.
\end{equation}
The quantities $\gamma_j(t)$ define real decay rates whose time dependence is generally unconstrained by the Hermiticity and trace preserving properties of $\Phi_t$. 
However, for a dynamical map which is CP-divisible, these decay rates must be nonnegative as proven by the following theorem.

\begin{theorem}[CP-divisibility = $\gamma_j(t)\geq0$]\label{theorem:5}
     Let $\Phi_t$ define an invertible map on $\mathcal{B}(\mathcal{H}_S)$ with $\Phi_0=\mathcal{I}_S$ and $d={\rm dim}\,\mathcal{H}_S$. 
     Then $\Phi_t$ is CP-divisible if and only if the time-local generator \eqref{eq:tlg} can be written in the canonical form \eqref{eq:canonical} with
    \begin{equation}
        \gamma_j(t)\geq 0, \qquad \forall j,
    \end{equation}
    for all $t\geq 0$.
\end{theorem}

A proof can be found in \cite{Rivas2010}. \qed 

\bigskip

\noindent A very important subclass of CP-divisible processes $\{\Phi_t = e^{\mathcal{L}t}\}_{t\geq 0}$ whose generator $\mathcal{L}$ is in the time independent form
\begin{equation}\label{eq:gksl}
    \mathcal{L}\rho = -i[H_0,\rho] + \sum^{d^2-1}_{j=1}\gamma_j\Big[L_j\rho L^{\dagger}_j - \tfrac{1}{2}L^{\dagger}_jL_j\rho - \tfrac{1}{2}\rho L^{\dagger}_jL_j\Big]
\end{equation}
define what are known as quantum dynamical semigroups \cite{BP2002,Alicki1987}. 
This name derives from the fact that such maps obey the semigroup property $\Phi_{t+s} = \Phi_t\Phi_s$, $t,s\geq0$, 
which follows as a consequence of $\Lambda_{t,s}$ belonging to the same one-parameter family as $\Phi_t=e^{\mathcal{L}t}$; i.e., $\Lambda_{t,s} = \Phi_t\Phi^{-1}_s = e^{\mathcal{L}(t-s)}$, and so $\Lambda_{t,s} = \Lambda_{t-s} = \Phi_{t-s}$.
In this regard, quantum dynamical semigroups describe time homogenous processes analogously to how Markov semigroups \eqref{eq:cke_semigroup} describe time homogenous Markov chains. At the same time, generators of the type \eqref{eq:gksl} are said to be in Gorini-Kossakowski-Sudarshan-Lindblad (GKSL) form as a result of the well known GKSL theorem \cite{Gorini1976,Lindblad1976}.
This theorem states that a family of CPTP maps $\{\Phi_t : t\geq0\}$ forms a quantum dynamical semigroup if and only if $\Phi_t = e^{\mathcal{L}t}$, 
where $\mathcal{L}$ can be written in the canonical form above with nonnegative decay rates $\gamma_j\geq0$.   
Accordingly, Theorem \ref{theorem:5} is essentially an extension of the GKSL theorem to time-dependent generators $\mathcal{L}_t$ for which $\Lambda_{t,s}\neq\Lambda_{t-s}$, and 
\begin{equation}\label{eq:cptp_tip}
    \Lambda_{t,s} = \mathcal{T}e^{\int^t_sdu\,\mathcal{L}_u}.
\end{equation}
Here, $\mathcal{T}$ is the so-called chronological time-ordering superoperator\footnote{For any two maps $\mathcal{L}_t$ and $\mathcal{L}_s$, the time-ordering superoperator $\mathcal{T}$ arranges them chronologically with decreasing time argument, $\mathcal{T}\mathcal{L}_t\mathcal{L}_s = \theta(t-s)\mathcal{L}_t\mathcal{L}_s + \theta(s-t)\mathcal{L}_s\mathcal{L}_t$, where $\theta(t)$ is the Heaviside-step function.} (i.e., a map on the space of bounded maps). 
It should also be noted that like their classical counterpart \eqref{eq:cke_semigroup}, quantum dynamical semigroups are intimately connected with memoryless processes \cite{BP2002}.  
For the sake of completeness, we include in Appendix \ref{appenA} a derivation showing how \eqref{eq:gksl} can be microscopically derived under the Born, secular, and Markov approximations, where the latter physically encodes the memoryless property of \eqref{eq:gksl}. This connection will also be further discussed in the next section. 

Finally, the weaker condition of P-divisibility is ensured through the generator $\mathcal{L}_t$ satisfying the following property \cite{Breuer2016}.

\begin{proposition}[P-divisibility = \eqref{eq:pdiv_gen}]
Let $\Phi_t$ define an invertible map on $\mathcal{B}(\mathcal{H}_S)$ with $\Phi_0=\mathcal{I}_S$ and $d={\rm dim}\,\mathcal{H}_S$. 
Then $\Phi_t$ is P-divisible if and only if the time-local generator \eqref{eq:tlg} can be written in the canonical form \eqref{eq:canonical} and
\begin{equation}\label{eq:pdiv_gen}
    \sum_j\gamma_j(t)|\langle n|L_j(t)|m\rangle|^2 \geq 0
\end{equation}
for $n\neq m$ and $t\geq0$, where $\{|n\rangle\}^{d}_{n=1}$ defines any orthonormal basis of $\mathcal{H}_S$. 
\end{proposition}

\noindent A corollary of this result is that for master equations with a single decay channel, 
the corresponding decay rate $\gamma(t)$ must be nonnegative for all $t\geq0$ provided the map $\Phi_t$ is P-divisible. 
In turn, this implies CP-divisibility and P-divisibility to be equivalent in such cases, since $\gamma(t)\geq0$ also guarantees the same map to be CP-divisible
(i.e., if $\Phi_t$ is P-divisible, then it is also CP-divisible, and vice versa).

\section{Quantum characterizations of Markovianity}\label{sec:4}

We now turn to discussing quantum notions of Markovianity in view of the formal classical definition \eqref{eq:Markov} based on past-state independence.
To first motivate this discussion, we outline why such a classical definition of Markovianity cannot be wholly extended to the quantum domain \cite{Breuer2016,Rivas2014,Li2018}.

\subsection{Violation of the Kolmogorov consistency conditions}

Consider a quantum system characterized by a complete set of commuting observables (CSCO) $\{O_1,...,O_m\}$, which is in state $\rho(t_1)$ at time $t_1$. 
The commutativity of the operators $[O_i,O_j]=0$ $\forall i,j$ implies the existence of common eigenbasis $\{|x_{t_1}\rangle\}$,
where $x_{t_1} = (o_1,...,o_m)$ labels the corresponding set of eigenvalues of the CSCO. 
According to Born's rule, the probability of measuring the system in state $x_{t_1}$ is 
\begin{equation}\label{eq:born}
    P_1(x_{t_1}) = {\rm Tr}[\Pi_{x_{t_1}}\rho(t_1)\Pi_{x_{t_1}}] = {\rm Tr}[\Pi_{x_{t_1}}\rho(t_1)],
\end{equation}
with $\Pi_{x_{t_1}} = |x_{t_1}\rangle\langle x_{t_1}|$. 
We highlight that this distribution is consistent with the one-point probabilities of a classical stochastic process provided $\rho(t_1)$ is also diagonal in the same eigenbasis as the CSCO.

Consider now the time evolution under a unitary map $\Phi_t\rho = U(t)\rho U^{\dagger}(t)$, starting from an initial state $\rho(t_1) = \sum_{x_{t_1}} P_1(x_{t_1}) \Pi_{x_{t_1}}$. 
If the system is projectively measured in the bases $\{|x_t\rangle : t\in T\}$ at times $t_n\geq...\geq t_1$, then the joint probability of measuring $x_{t_n},...,x_{t_1}$ is given by 
\begin{equation}\label{eq:joint_born}
    P_n(x_{t_n},...,x_{t_1}) = {\rm Tr}[\mathcal{P}_{x_{t_n}}\Phi_{t_n-t_{n-1}}\mathcal{P}_{x_{t_{n-1}}}...\,\Phi_{t_2-t_1}\mathcal{P}_{x_{t_1}}\rho(t_1)],
\end{equation}
where $\mathcal{P}_x\rho = \Pi_x\rho\Pi_x$. 
Based on completeness of the projectors $\{\Pi_{x_t}\}_{t\geq 0}$, it is straightforward to verify that the joint distributions $P_n(x_{t_n},...,x_{t_1})$ are suitably normalized, and hence may be interpreted as classical probabilities akin to \eqref{eq:born} (as per Born's rule). 
This naturally leads us to ask: what differentiates this family of joint distributions from those used to describe classical stochastic processes?
The answer is that they generally violate the Kolmogorov consistency conditions \cite{Breuer2016,Li2018,Vacchini2011}.
In particular, at intermediate times $t_n>t_j>t_1$, we have 
\begin{align}\label{eq:kcc_violation}
    \sum_{x_{t_j}}P_n(x_{t_n},...,x_{t_1}) &= {\rm Tr}[\mathcal{P}_{x_{t_n}}\Phi_{t_n-t_{n-1}}...\sum_{x_{t_j}}\mathcal{P}_{x_{t_j}}\,\Phi_{t_2-t_1}\mathcal{P}_{x_{t_1}}\rho(t_1)] \nonumber\\
                                           &\neq P_{n-1}(x_{t_n},...,\cancel{x_{t_j}},...,x_{t_1}),
\end{align}
such that the $n$-point distributions \eqref{eq:joint_born} do not form a connected hierarchy like in the classical case.
An exception to this is when the basis $\{|x_t\rangle\}$ is time invariant,
so that the state $\rho(t_1)$ remains diagonal in the initial eigenbasis. 
Under this constraint, all the operators inside the trace commute leading to \eqref{eq:kcc1}-\eqref{eq:kcc2} being satisfied. 
This indicates how the violation of the Kolmogorov consistency conditions is precisely due to the initial CSCO changing at later times $t>t_1$,
leading in turn to the generation of coherences in the eigenbasis $\{|x_{t_1}\rangle\}$\footnote{Due to these coherences, performing measurements will generally disturb the system, thereby affecting its posterior statistics---see \cite{Milz2020}.}. 
Obviously, when no such coherences are generated, the system effectively behaves classically \cite{Milz2020}.
A key consequence of \eqref{eq:kcc_violation} therefore is that the classical notion of Markovianity is ambiguous for quantum processes, 
since quantum conditional probabilities cannot be defined in the same way as \eqref{eq:st_cond_prob} via Bayes' rule.
It should be noted, however, that certain operational definitions of QM are known to recover the Markov property \eqref{eq:Markov} within a suitable classical limit \cite{Pollock2018b}.

\subsection{Definitions based on the dynamical map}

In lieu of a single fixed definition of quantum Markovianity (QM), various QM-related concepts have been explored in literature \cite{Li2018}.
These concepts broadly align with two distinct approaches to characterizing memory effects in open quantum system dynamics: (i) the intrinsic approach, which focuses solely on the properties of the dynamical map and/or its local effect on the reduced system entangled with an ancilla \cite{Rivas2010b, Luo2012}; 
and (ii) the extrinsic approach, focusing on properties of both the system and its environment \cite{Pollock2018a,Pollock2018b,Gullo2014}. 
For simplicity, we shall only consider QM definitions aligned with the first approach, 
which provide a weaker characterization of Markovianity compared to the second. 
A more comprehensive overview of these concepts and their relation to classical Markovianity measures can be found in \cite{Li2018}.

\subsubsection{CP-divisibility}

The first characterization of QM we consider is that based on the CP-divisibility of the dynamical map $\Phi_t$.

\begin{definition}[CP-divisibility = Markovian]\label{def:6}
    
A quantum process $\{\Phi_t : t\geq 0\}$ is Markovian if for all $t\geq s \geq 0$, the map $\Phi_t$ is CP-divisible according to Definition \ref{def:3}.
Equivalently, it is Markovian if the corresponding canonical-form master equation \eqref{eq:canonical} only has nonnegative decay rates.

\end{definition}

\medskip 

\noindent In essence, this extends the P-divisibility criterion for classical Markov processes to quantum dynamical maps.
It was originally proposed by Wolf and Cirac \cite{Wolf2008a,Wolf2008b} in the context of quantum dynamical semigroups, 
and later generalized by Rivas, Huelga, and Plenio \cite{Rivas2010b} to time inhomogenous processes whereby $\Lambda_{t,s}\neq \Lambda_{t-s}$. 
The same authors also used this definition to construct a quantifiable measure of non-Markovianity through the Choi-Jamiołkowski isomorphism \cite{Choi1972}. 
In particular, let
\begin{equation}\label{eq:g}
    g(t) := \lim_{\epsilon\to 0^+}\frac{\|C_{\Lambda}(t+\epsilon, t)\|_1 - 1}{\epsilon}
\end{equation}
where $C_{\Lambda}(t+\epsilon, t) = (\Lambda_{t+\epsilon,t} \otimes \mathcal{I}^{(d)}_A)|\Phi^+_d\rangle\langle\Phi^+_d|$ is the the Choi matrix of the map $\Lambda_{t+\epsilon,t}$ (see \eqref{eq:choi}).
It can then be shown that $\|C_{\Lambda}(t+\epsilon, t)\|_1 = 1$ holds if and only if $\Lambda_{t+\epsilon,t}$ is CPTP through the following result.

\begin{proposition}[CPTP $\Lambda$ $\Leftrightarrow$ $\|C_{\Lambda}\|_1 = 1$]
    Let $C_{\Lambda}$ be the Choi matrix of a map $\Lambda$ on $\mathcal{B}(\mathcal{H}_S)$ with $d={\rm dim}\,\mathcal{H}_S$. 
     Then $\Lambda$ is CPTP if and only if $\|C_{\Lambda}\|_1 = 1$, and $\|C_{\Lambda}\|_1 > 1$ otherwise.
\end{proposition}

A proof is provided in Appendix \ref{appenB}. \qed

\bigskip

\noindent Hence, the measure 
\begin{equation}\label{eq:RHP}
    \mathcal{N}_{\rm RHP}(\Phi) = \int^{\infty}_0dt\,g(t)
\end{equation}
will be nonzero if and only if the map $\Phi_t$ is non-Markovian according to Definition \ref{def:6}. 

It is evident that the above definition of $g(t)$ applies to continuous time processes. 
However, an equivalent measure applying to discrete time processes $\{\Phi_t : t=n\delta t, n\in\mathbb{N}\}$ may be defined from
\begin{equation}
    g_t := \frac{\|C_{t + \delta t, t}\|_1 - 1}{\delta t}, \qquad t=n\delta t,
\end{equation}
where $\mathcal{N}_{\rm RHP}(\Phi) = \sum_{t} g_t$. 
Naturally, this measure converges to its continuous time counterpart in the limit $\delta t=t/n\rightarrow 0^+$.

\subsubsection{Monotonically decreasing state distinguishability (MDSD)}

An alternative characterization of QM proposed by Breuer, Laine, and Piilo \cite{Breuer2016,Breuer2009,Laine2010} is related to the monotonic decay of the trace distance \eqref{eq:td}, mirroring the property of decreasing Kolmogorov distance for P-divisible classical processes.

\begin{definition}[MDSD = Markovian]\label{def:7}
    
A quantum process $\{\Phi_t : t\geq 0\}$ is Markovian if for all $t\geq s \geq 0$, 
the trace distance between any pair of initial states $\{\rho^1_S,\rho^2_S\}\in\mathcal{S}(\mathcal{H}_S)$ is monotonically decreasing under the action of $\Phi_t$: namely,
\begin{equation}\label{eq:mdsd}
    D(\Phi_t(\rho^1_S),\Phi_t(\rho^2_S)) \leq D(\Phi_s(\rho^1_S),\Phi_s(\rho^2_S)). 
\end{equation}

\end{definition}

\medskip

\noindent It is worth noting like \eqref{eq:kd_contract} that nonmonotonic increases in the trace distance do not contradict the contractivity property \eqref{eq:td_cptp}.
The MDSD condition also exhibits a clear information theoretic interpretation based on the idea of information flows between the system and environment (see Sec. \ref{sec:trace_dist} for a discussion on the trace distance serving as an information quantifier). 
In particular, QM according to Definition \ref{def:7} is associated with a continuous loss of information from the system to the environment, 
represented by a monotonically decreasing state distinguishability,
while non-Markovian behavior is signified by information backflow where the state distinguishability temporarily increases \cite{Breuer2016,Vacchini2024}.
This same definition implies that the quantity 
\begin{equation}\label{eq:sigma}
    \sigma(\rho^{1,2}_S,t) := \frac{d}{dt}D(\Phi_t(\rho^1_S),\Phi_t(\rho^2_S))
\end{equation}
will be nonnegative if an only if \eqref{eq:mdsd} is violated for some $t\geq 0$.

The following proposition also proves that MDSD and $\sigma(\rho^{1,2}_S,t)<0$ coincide with the map $\Phi_t$ being P-divisible \cite{Chruscinski2022}. 

\begin{proposition}
    Let $\Phi_t$ define a map on $\mathcal{B}(\mathcal{H}_S)$ for $t\geq 0$. 
    Then $\Phi_t$ is P-divisible if and only if 
    \begin{equation}\label{eq:trace_norm_contract}
        \frac{d}{dt}\|\Phi_t(X)\|_1 \leq 0,
    \end{equation}
    for any Hermitian operator $X\in\mathcal{B}(\mathcal{H}_S)$. 
\end{proposition} 
Provided $\Phi_t$ is invertible, we can write 
\begin{align*}
     \frac{d}{dt}\|\Phi_t(X)\|_1 &= \lim_{\epsilon\rightarrow 0^+}\frac{\|\Phi_{t+\epsilon}(X)\|_1 - \|\Phi_t(X)\|_1}{\epsilon} \\
                                 &= \lim_{\epsilon\rightarrow 0^+}\frac{\|\Lambda_{t+\epsilon,t}\Phi_t(X)\|_1 - \|\Phi_t(X)\|_1}{\epsilon} \\
                                 &= \lim_{\epsilon\rightarrow 0^+}\frac{\|\Lambda_{t+\epsilon,t}(Y)\|_1 - \|Y\|_1}{\epsilon},
\end{align*}
where $Y \in\mathcal{B}(\mathcal{H}_S)$ and $Y=Y^{\dagger}$. 
We then employ the spectral decomposition $Y = Y^+ - Y^-$ (with $Y^{\pm}$ representing the positive and negative parts of the spectrum of $Y$) to obtain
\begin{align*}
    \|\Lambda_{t+\epsilon,t}(Y)\|_1 &= {\rm Tr}|\Lambda_{t+\epsilon,t}(Y^+ - Y^-)| \\
                                    &\leq {\rm Tr}|\Lambda_{t+\epsilon,t}Y^+| + {\rm Tr}|\Lambda_{t+\epsilon,t}Y^-| \hspace{-2cm} &&  (\text{Triangle inequality}) \\
                                    &= {\rm Tr}(\Lambda_{t+\epsilon,t} Y^+) + {\rm Tr}(\Lambda_{t+\epsilon,t} Y^-) \hspace{-2cm} &&  (\text{Positivity}) \\
                                    &= {\rm Tr}(Y^+) + {\rm Tr}(Y^-). \hspace{-2cm} &&  (\text{Trace preservation})
\end{align*}
Likewise, the trace norm of $Y$ is given by 
\begin{equation*}
    \|Y\|_1 = {\rm Tr}|Y| = {\rm Tr}(Y^+) + {\rm Tr}(Y^-),
\end{equation*}
such that $\|\Lambda_{t+\epsilon,t}(Y)\|_1\leq \|Y\|_1$, from which \eqref{eq:trace_norm_contract} follows. 
Conversely, assume \eqref{eq:trace_norm_contract} holds. Then due to the contractivity of positive maps $\|\Lambda_{t+\epsilon,t}(Y)\|_1\leq \|Y\|_1$ \cite{Paulsen2003},
this infers $\Lambda_{t+\epsilon,t}$ must be positive for $t\geq 0$, so that $\Phi_t$ is P-divisible. \qed 

\medskip

\medskip

\noindent Accordingly, one can define the measure \cite{Breuer2009}
\begin{equation}\label{eq:BLP}
    \mathcal{N}_{\rm BLP}(\Phi) = \max_{\rho^{1,2}_S}\int_{\sigma>0} dt\,\sigma(\rho^{1,2}_S,t), 
\end{equation}
which is zero if and only if the intermediate map $\Lambda_{t,s}$ is P-divisible for $t\geq s\geq 0$. 
Note that unlike \eqref{eq:RHP}, the measure $\mathcal{N}_{\rm BLP}(\Phi)$ requires optimization over all pairs of possible input states $\{\rho^1_S,\rho^2_S\}$. 

Recall from the definition of complete positivity that all CPTP maps are PTP. 
Thus, we have the following relationship between $\mathcal{N}_{\rm RHP}(\Phi)$ and $\mathcal{N}_{\rm BLP}(\Phi)$ \cite{Chruscinski2011}.

\begin{theorem}[CP-divisibility $\Rightarrow$ MDSD]

    If a quantum process $\{\Phi_t : t\geq 0\}$ is Markovian according to Definition \ref{def:6}, then it is also Markovian according to Definition \ref{def:7}, but not vice versa:
    \begin{equation}
        \mathcal{N}_{\rm RHP}(\Phi)=0 \Rightarrow \mathcal{N}_{\rm BLP}(\Phi) = 0.
    \end{equation}
    Equivalently, $\mathcal{N}_{\rm BLP}(\Phi)>0$ defines a necessary condition for the CP-divisibility of the dynamics. 

\end{theorem}

\medskip

\noindent A special case where these measures formally coincide is for quantum dynamical semigroups with a corresponding GKSL form generator \eqref{eq:gksl}.

\begin{corollary}[Quantum dynamical semigroups = Markovian]\label{cor:1}
    Since quantum dynamical semigroups $\{\Phi_t = e^{\mathcal{L}t}\}_{t\geq 0}$ are CP-divisible by construction, both the RHP and BLP measures predict semigroup processes to be Markovian, i.e., $\mathcal{N}_{\rm BLP}(\Phi) = \mathcal{N}_{\rm RHP}(\Phi) = 0$ for all $t\geq 0$. 
\end{corollary}

\bigskip 

\noindent Hence, quantum dynamical semigroups have a clear association with time homogenous Markov processes like with Markov semigroups in the classical regime.

Finally, an analogous discrete time form of the quantity $\sigma(\rho^{1,2}_S,t)$ can be defined as 
\begin{equation}
    \sigma_t(\rho^{1,2}_S) = \frac{D(\Phi_{t+\delta t}(\rho^1_S), \Phi_{t+\delta t}(\rho^2_S)) - D(\Phi_{t}(\rho^1_S), \Phi_{t}(\rho^2_S))}{\delta t},
\end{equation}
which converges to \eqref{eq:sigma} in the limit $\delta t = t/n \rightarrow 0$.

\section{Example: spin-boson model}\label{sec:5}

Let us now consider an application of the non-Markovianity measures \eqref{eq:RHP} and \eqref{eq:BLP} to a paradigmatic model of open quantum systems---the spin-boson model \cite{Leggett1987,Stockburger2004}.
This model describes a spin-$\frac{1}{2}$ or qubit system $S$ interacting with a bath of bosonic particles $B$ represented by non-interacting harmonic oscillators.  
Accordingly, the system state space is $\mathcal{H}_S={\rm span}\{|0\rangle,|1\rangle\}=\mathbb{C}^2$, where $|0\rangle$ ($|1\rangle$) denotes the ground (excited) state of the qubit.
The model can physically describe dissipation and decoherence in a wide variety of quantum systems whose state space may be approximately restricted to the lowest energy manifold (or exactly in the case of a spin-$\frac{1}{2}$ particle), including superconducting qubits \cite{Weiss2012}, atom-cavity systems, and trapped ions \cite{Porras2008}. In the following, we consider the simplest formulation in which there is no direct coupling between the states $|0\rangle$ and $|1\rangle$, and where the rotating wave approximation is applied to the qubit-bath interaction (see below). Note that the non-Markovianity for other formulations of the spin-boson model has been analyzed in \cite{Clos2012,Haikka2013,Bylicka2014,Addis2014,Guarnieri2016} (see also \cite{Einsiedler2020} for the related Caldeira-Leggett model).

The free Hamiltonians of the qubit and bath are 
\begin{equation}
    H_S = \omega_0\sigma_+\sigma_-, \qquad H_B = \sum_k\omega_k b^{\dagger}_kb_k.
\end{equation}
Here, $\sigma_+=|1\rangle\langle 0|$ ($\sigma_- = |0\rangle\langle 1|$) is the Pauli raising (lowering) operator, 
$\omega_0$ is the qubit energy splitting,
and $b_k$ ($b^{\dagger}_k$) is the annihilation (creation) operator of the $k$-th bath mode with frequency $\omega_k$. Within the rotating wave approximation \cite{Tannoudji1998}, the interaction Hamiltonian reads
\begin{equation}\label{eq:H_int}
    H_I = \sum_k(g_k\sigma_+\otimes b_k + g^*_k\sigma_-\otimes b^{\dagger}_k),
\end{equation}
where $g_k$ denotes the coupling strength of the $k$-th bath mode to the qubit. 
Further details on this approximation and methods to solve the resulting dynamics can be found in \cite{BP2002}.

A crucial aspect of this model is that it is exactly solvable by virtue of the Hamiltonian \eqref{eq:H} conserving the total excitation number \cite{Garraway1997a,Garraway1997b}.
In particular, with the bath initially in the vacuum state $\rho_B(0) = |\boldsymbol{0}\rangle\langle \boldsymbol{0}|_B$ ($|\boldsymbol{0}\rangle_B := |0_k\rangle^{\otimes k}_B$, $b_k|\boldsymbol{0}\rangle_B=0$), 
the reduced system state $\rho_S(t)$ evolves in the interaction picture as \cite{BP2002,Merkli2026}
\begin{align}\label{eq:map_sb}
    \rho_S(0) \mapsto \Phi_t\rho_S(0) &=
    \rho_{11}|G(t)|^2|1\rangle\langle 1| + \rho_{10}G(t)|1\rangle\langle 0| \nonumber\\
    &+ \rho^*_{10}G^*(t)|0\rangle\langle 1| + (1 - \rho_{11}|G(t)|^2)|0\rangle\langle 0|,
\end{align}
where $G(t)$ obeys the integro-differential equation
\begin{equation}\label{eq:G_eom}
    \frac{d}{dt}G(t) = -\int^t_0 ds\,f(t-s)G(s), \qquad G(0) = 1,
\end{equation}
and $f(t-s)$ is the bath correlation function. 
This correlation function may be expressed in terms of the spectral density $J(\omega) = \sum_k|g_k|^2\delta(\omega-\omega_k)$ as 
\begin{equation}\label{eq:bcf_sb}
    f(t-s) = \int^{\infty}_{-\infty} d\omega\,J(\omega)e^{-i(\omega-\omega_0)(t-s)}.
\end{equation}
Physically, $J(\omega)$ measures the coupling response of the bath modes to the system as a function of mode frequency $\omega$. 
To illustrate how the kernel \eqref{eq:bcf_sb} is connected to the bath memory,
we may consider a flat spectral density $J(\omega)\sim \gamma_0$ corresponding to when the bath modes couple uniformly (weakly) to the qubit.
Since the correlation function in this limit becomes singular $f(t-s)\sim \gamma_0\delta(t-s)$,
it is evident from \eqref{eq:G_eom} that the qubit population and coherences undergo exponential decay at a rate $\gamma_0$, i.e., $G(t) = \exp(-\gamma_0 t/2)$. 
It is also possible to show from \eqref{eq:G_eom} that the reduced system density matrix obeys the GKSL master equation (i.e., a canonical form master equation with generator \eqref{eq:gksl})
\begin{equation}\label{eq:sb_gksl}
    \frac{d}{dt}\rho_S(t) = \mathcal{L}\rho_S(t) = \gamma_0\Big[\sigma_-\rho_S(t)\sigma_+ - \tfrac{1}{2}\sigma_+\sigma_-\rho_S(t) - \tfrac{1}{2}\rho_S(t)\sigma_+\sigma_-\Big].
\end{equation}
Hence, the dynamics is Markovian according to both the CP-divisibility and MDSD criteria (see Corollary \ref{cor:1}). 
We note that the association of QM with a flat spectral density $J(\omega)$ is consistent with the limit where the Born-Markov approximations are valid---see Appendix \ref{appenA} \cite{Tannoudji1998}. 

Let us now consider the most general differential form of the map \eqref{eq:map_sb}. 
Following \cite{Merkli2026}, the reduced density matrix $\rho_S(t)$ can be shown to obey the exact time-local master equation \cite{BP2002}
\begin{align}\label{eq:sb_tcl}
    \frac{d}{dt}\rho_S(t) &= \mathcal{L}_t\rho_S(t) \nonumber\\
                          &= -\tfrac{i}{2}[S(t)\sigma_+\sigma_-,\rho_S(t)] + \gamma(t)\Big[\sigma_-\rho_S(t)\sigma_+ - \tfrac{1}{2}\{\sigma_+\sigma_-,\rho_S(t)\}\Big],
\end{align}
where the time-dependent Lamb shift $S(t)$ and decay rate $\gamma(t)$ are given by 
\begin{equation}\label{eq:S_gamma}
    S(t) = -2{\rm Im}\,\bigg[\frac{\dot{G}(t)}{G(t)}\bigg], \quad \gamma(t) = -2{\rm Re}\,\bigg[\frac{\dot{G}(t)}{G(t)}\bigg]. 
\end{equation}
This master equation gives rise to the following equations of motion for the matrix elements $\dot{\rho}_{ij}(t) = \langle i|\mathcal{L}_t\rho_S(t)|j\rangle$, $i,j\in\{0,1\}$,
\begin{align}
    \dot{\rho}_{11}(t) &= -\gamma(t)\rho_{11}(t), \hspace{-2cm} &&\rho_{00}(t) = 1 - \rho_{11}(t),\\
    \dot{\rho}_{10}(t) &= -\tfrac{1}{2}[iS(t) + \gamma(t)]\rho_{10}(t), \hspace{-2cm} &&\rho_{01}(t) = \rho^*_{10}(t),
\end{align}
which can easily be solved to obtain (compare with \eqref{eq:S_gamma})
\begin{equation}\label{eq:G_exact}
    G(t)= \exp\bigg(-\frac{1}{2}\int^t_0 ds\,[iS(s) + \gamma(s)]\bigg).
\end{equation}
Accordingly, it can be found that a necessary and sufficient condition for $\Phi_t$ to be CPTP is $ \int^t_0 ds\,\gamma(s) \geq 0$ (see Exercise 3).
The time-local master equation may also be used to evaluate the non-Markovianity measures $\mathcal{N}_{\rm RHP}(\Phi)$ and $\mathcal{N}_{\rm BLP}(\Phi)$, as we will now show.

\subsection{Evaluation of $\mathcal{N}_{\rm RHP}$}\label{sec:RHP}

Using the expansion $\Lambda_{t+\epsilon,t} = \mathcal{I}_S + \mathcal{L}_t\epsilon + O(\epsilon^2)$ (see \eqref{eq:cptp_tip}),
the quantity $g(t)$ may first be simplified to \cite{Rivas2014,Rivas2010b}
\begin{equation}\label{eq:g_tl}
    g(t) = \lim_{\epsilon\rightarrow 0^+}\frac{\|(\mathcal{I}_{SA} + \epsilon\mathcal{L}_t\otimes\mathcal{I}^{(2)}_A)|\Phi^+_2\rangle\langle 
        \Phi^+_2|\|_1 - 1}{\epsilon},
\end{equation}
having used that terms beyond first order in $\epsilon$ vanish in the limit $\epsilon\rightarrow0^+$. 
The next step is to represent the state $|\Phi^+_2\rangle\langle \Phi^+_2|$ 
using a suitable orthonormal basis for $\mathcal{B}(\mathbb{C}^2\otimes\mathbb{C}^2)$. 
Since this operator space admits a tensor product structure $\mathcal{B}(\mathbb{C}^2\otimes\mathbb{C}^2) = \mathcal{B}(\mathbb{C}^2)\otimes\mathcal{B}(\mathbb{C}^2)$,
we may choose the Pauli basis $\mathcal{B}(\mathbb{C}^2\otimes \mathbb{C}^2) = \{\sigma_i\otimes\sigma_j\}^3_{i,j=0}$
\begin{equation}\label{eq:hs_basis}
    \sigma_0 = \frac{1}{\sqrt{2}}\mathbb{I}_S, \quad \sigma_1 = \sigma_+, \quad \sigma_2 = \sigma_-, \quad \sigma_3 = \frac{1}{\sqrt{2}}\sigma_z,
\end{equation}
where the operators $\{\sigma_i\}^3_{i=0}\in\mathcal{B}(\mathbb{C}^2)$ are orthonormal with respect to the Hilbert-Schmidt inner product 
\begin{equation}
    (\sigma_i,\sigma_j) := {\rm Tr}(\sigma^{\dagger}_i\sigma_j) = \delta_{ij}.
\end{equation}
Accordingly, we have $|\Phi^+_2\rangle\langle \Phi^+_2| = \sum_{i,j}c_{ij}\sigma_i\otimes\sigma_j$, and
\begin{align}
    c_{ij} &:= {\rm Tr}[(\sigma_i\otimes\sigma_j)^{\dagger}|\Phi^+_2\rangle\langle \Phi^+_2|] \nonumber \\
           &= \frac{1}{2}\sum_{n,m}(\sigma_i,|n\rangle\langle m|) \, (\sigma_j,|n\rangle\langle m|).
\end{align}
The only nonzero coefficients $c_{ij}$ in the basis \eqref{eq:hs_basis} are found to be
\begin{equation}
    c_{00} = c_{11} = c_{22} = c_{33} = \frac{1}{2}. 
\end{equation}
It now suffices to apply the time-local generator $\mathcal{L}_t$ of \eqref{eq:sb_tcl} to each of the terms in the expansion of $|\Phi^+_2\rangle\langle \Phi^+_2|$; that is, we want to evaluate    
\begin{equation}
    (\mathcal{L}_t\otimes\mathcal{I}^{(2)}_A)|\Phi^+_2\rangle\langle \Phi^+_2| = \frac{1}{2}\sum_i\mathcal{L}_t[\sigma_i]\otimes\sigma_i.
\end{equation}
To do so, one can make use of the fact that the operators $\{\sigma_j\}_j$ are closed under the action of $\mathcal{L}_t$: 
\begin{align}
\begin{split}
    \mathcal{L}_t[\sigma_0] &= -\gamma(t)\sigma_3, \\ 
    \mathcal{L}_t[\sigma_1] &= \tfrac{1}{2}\big[iS(t) - \gamma(t)]\sigma_1, \\
    \mathcal{L}_t[\sigma_2] &= -\tfrac{1}{2}\big[iS(t) + \gamma(t)]\sigma_2, \\
    \mathcal{L}_t[\sigma_3] &= -\gamma(t)\sigma_3.
\end{split}
\end{align}
Hence, we obtain
\begin{equation}
    (\mathcal{L}_t\otimes\mathcal{I}^{(2)}_A)|\Phi^+_2\rangle\langle \Phi^+_2| 
    = -\frac{1}{2}\Big[\gamma(t)\sigma_3\otimes\sigma_0 + \tfrac{1}{2}\big[-iS(t) + \gamma(t)]\sigma_1\otimes\sigma_1 
    + \tfrac{1}{2}\big[iS(t) + \gamma(t)]\sigma_2\otimes\sigma_2 + \gamma(t)\sigma_3\otimes\sigma_3\Big],
\end{equation}
yielding the Choi matrix
\begin{align}\label{eq:choi_mat}
    (\mathcal{I}_{SA} + \epsilon\mathcal{L}_t\otimes\mathcal{I}^{(2)}_A)|\Phi^+_2\rangle\langle \Phi^+_2|
    &= \frac{1}{4}\mathbb{I}\otimes\mathbb{I} 
    - \tfrac{1}{4}\epsilon\gamma(t)\sigma_z\otimes\mathbb{I} 
    + \tfrac{1}{2}(1 + \tfrac{1}{2}[iS(t) - \gamma(t)]\epsilon)\sigma_-\otimes\sigma_-  \nonumber\\
    &+ \tfrac{1}{2}(1 - \tfrac{1}{2}[iS(t) + \gamma(t)]\epsilon)\sigma_+\otimes\sigma_+ + \tfrac{1}{4}(1-\gamma(t)\epsilon)\sigma_z\otimes\sigma_z.
\end{align}
This matrix can be expressed in the ordered basis $\{|11\rangle,|10\rangle,|01\rangle,|00\rangle\}$ as 
\begin{equation}
     (\mathcal{I}_{SA} + \epsilon\mathcal{L}_t\otimes\mathcal{I}^{(2)}_A)|\Phi^+_2\rangle\langle \Phi^+_2| =
     \begin{pmatrix}
        \tfrac{1}{2}(1 - \gamma(t)\epsilon) & 0 & 0 & \tfrac{1}{2}(1 - \tfrac{1}{2}[iS(t) + \gamma(t)]\epsilon) \\
        0 & 0 & 0 & 0 \\
        0 & 0 & \tfrac{1}{2}\gamma(t)\epsilon & 0 \\
        \tfrac{1}{2}(1 + \tfrac{1}{2}[iS(t) - \gamma(t)]\epsilon) & 0 & 0 & \tfrac{1}{2}
     \end{pmatrix}
\end{equation}
such that the trace norm of \eqref{eq:choi_mat} can be evaluated as
$\|(\mathcal{I}_{SA} + \epsilon\mathcal{L}_t\otimes\mathcal{I}^{(2)}_A)|\Phi^+_2\rangle\langle \Phi^+_2|\|_1 = \tfrac{1}{2}|1 - \gamma(t)\epsilon| + \tfrac{1}{2}|\gamma(t)|\epsilon + \tfrac{1}{2}$.
Finally, we may subsitute the above into \eqref{eq:g_tl}
\begin{align}
    g(t) &= \lim_{\epsilon\rightarrow 0^+}\frac{\tfrac{1}{2}|1-\gamma(t)\epsilon| + \tfrac{1}{2}|\gamma(t)|\epsilon + \tfrac{1}{2} - 1}{\epsilon} \nonumber\\
         &= \lim_{\epsilon\rightarrow 0^+}\frac{1 - \tfrac{1}{2}\gamma(t)\epsilon + \tfrac{1}{2}|\gamma(t)|\epsilon -1 + O(\epsilon^2)}{\epsilon} \nonumber\\
         &= \tfrac{1}{2}[|\gamma(t)| - \gamma(t)],
\end{align}
to obtain the measure \cite{Rivas2014}
\begin{equation}\label{eq:sb_RHP}
    \mathcal{N}_{\rm RHP}(\Phi) = 
    \begin{cases}
        0 & \text{if $\gamma(t)\geq 0$}, \\
        \int_{\gamma<0}dt\,|\gamma(t)| & \text{if $\gamma(t) < 0$}.
    \end{cases}
\end{equation}
This tells us that the open system dynamics will be non-Markovian provided the decay rate $\gamma(t)$ becomes negative during some (disjoint) time interval. 
This result is also consistent with Corollary \ref{cor:1}, since the corresponding GKSL master equation \eqref{eq:sb_gksl} always has a nonnegative decay rate $\gamma_0\geq 0$, and hence the dynamics is Markovian.

\subsection{Evaluation of $\mathcal{N}_{\rm BLP}$}

We start from the general expression for the trace distance (see \eqref{eq:td})
\begin{equation}\label{eq:td_qubit}
    D(\rho^1_S(t), \rho^2_S(t)) := \frac{1}{2}\|\Delta\rho_S(t)\|_1 = \frac{1}{2}|\lambda_-| + \frac{1}{2}|\lambda_+|, 
\end{equation}
where $\lambda_{\pm}$ denote the eigenvalues of $\Delta\rho_S(t) = \rho^1_S(t) - \rho^2_S(t)$. 
Note from \eqref{eq:map_sb} that any pair of open system states $\{\rho^1_S(t), \rho^2_S(t)\}$ can be written in the computational basis $\{|1\rangle, |0\rangle\}$ as 
\begin{equation}\label{eq:td_state_pair}
    \rho^{1,2}_S(t) = 
    \begin{pmatrix}
        |G(t)|^2\rho^{1,2}_{11} & G(t)\rho^{1,2}_{10} \\
        G^*(t)\rho^{1,2 *}_{10} & 1- |G(t)|^2\rho^{1,2}_{11}
    \end{pmatrix}.
\end{equation}
Using this expression, it is left as an exercise to the reader to show that (see Exercise 4)
\begin{equation}\label{eq:td_sb}
    \frac{1}{2}\|\Delta\rho_S(t)\|_1 = |G(t)|\sqrt{|G(t)|^2\Delta\rho^2_{11} + |\Delta\rho_{10}|^2},
\end{equation}
from which the time derivative is found to be
\begin{equation}\label{eq:td_sb_deriv}
    \sigma(\rho^{1,2}_S,t) = \frac{\frac{d}{dt}|G(t)|\,(2|G(t)|^2\Delta\rho^2_{11} + |\Delta\rho_{10}|^2)}{\sqrt{|G(t)|^2\Delta\rho^2_{11} + |\Delta\rho_{10}|^2}}.
\end{equation}
Since the sign of $\sigma(\rho^{1,2}_S,t)$ is controlled by $\frac{d}{dt}|G(t)|$, memory effects therefore emerge according to Definition \ref{def:7} only when $\frac{d}{dt}|G(t)|>0$.  
It is straightforward to determine that this condition is equivalent to a negative decay rate $\gamma(t)$: 
\begin{equation}
    \frac{d}{dt}|G(t)| > 0 \,\, \Leftrightarrow \,\, \gamma(t)<0.
\end{equation}
Moreover, it was shown in \cite{Wissmann2012} that the measure $\mathcal{N}_{\rm BLP}(\Phi)$ for a two-level system will be maximized for any pair of initial states $\{\rho^1_S,\rho^2_S\}$ with orthogonal support; namely,
those states which represent antipodal points on the Bloch sphere. 
Hence, without loss to generality we may evaluate $\mathcal{N}_{\rm BLP}(\Phi) = \max_{\rho^{1,2}_S}\int_{\sigma>0} dt\,\sigma(\rho^{1,2}_S,t)$ by choosing $\Delta\rho_{11} = 1$, $\Delta\rho_{10}=0$, such that 
\begin{equation}\label{eq:sb_BLP}
    \mathcal{N}_{\rm BLP}(\Phi) = 
     \begin{cases}
        0 & \text{if $\gamma(t)\geq 0$}, \\
        \sum_n\big(|G(t^f_n)|^2 - |G(t^i_n)|^2\big) & \text{if $\gamma(t) < 0$},
    \end{cases}
\end{equation}
where $t^i_n$ ($t^f_i$) denotes the start (end) point of the $n$-th time interval for which $\gamma(t)<0$. 

A direct comparison of \eqref{eq:sb_RHP} and \eqref{eq:sb_BLP} indicates that the RHP and BLP measures make identical predictions of non-Markovianity for this model \cite{Zeng2011}.
Notably, this agrees with our previous analysis that CP-divisibility and P-divisibility coincide for canonical-form master equations with a single decay channel (see Sec. \ref{sec:3.1.2}).

\subsection{Atom-cavity dynamics: Lorentzian spectral density}

To make our example more explicit, we may examine a commonly encountered form spectral density, namely a Lorentizan (see also \cite{BP2002,Merkli2026})
\begin{equation}
    J(\omega) = \frac{g^2}{\pi}\frac{\Gamma/2}{(\omega - \omega_0)^2 + (\Gamma/2)^2}.
\end{equation}
This spectral density describes the same reduced dynamics as when $S$ is a two-level atom coupled to a single cavity mode with strength $g$, and where the cavity mode undergoes Markovian dissipation at a rate $\Gamma$ \cite{Garraway1997b}. 
The corresponding bath correlation function \eqref{eq:bcf_sb} can be analytically determined as 
\begin{equation}\label{eq:bcf_lorz}
    f(t-s) = g^2 e^{-\Gamma|t-s|/2}, 
\end{equation}
from which \eqref{eq:G_eom} can be solved via Laplace transforms or the pseudomode method \cite{Garraway1997a,Pleasance2020} to obtain
\begin{equation}\label{eq:G_sol}
    G(t) = e^{-\Gamma t/4}\Big[\cos(\Omega t) + \frac{\Gamma}{4\Omega}\sin(\Omega t)\Big],
\end{equation}
with $\Omega = \tfrac{1}{2}\sqrt{4g^2 - (\Gamma/2)^2}$. 
From \eqref{eq:G_sol}, the dynamics can be seen to be non-Markovian depending on whether $\Omega$ is real of pure imaginary:
if $4g>\Gamma$, then $\Omega\in\mathbb{R}$, and the function $G(t)$ will show coherent oscillations. 
However, if $4g<\Gamma$, the propagator $G(t)$ will instead decay monotonically
given the sine and cosine terms become hyperbolic functions. 

The time-dependent Lamb shift $S(t)$ and decay rate $\gamma(t)$ can be determined from \eqref{eq:G_sol} as $S(t) = 0$ and 
\begin{equation}\label{eq:gm}
    \gamma(t) = \frac{2g^2\sin(\Omega t)}{\Omega\cos(\Omega t) + \tfrac{\Gamma}{4}\sin(\Omega t)}.
\end{equation}
Using this form of $\gamma(t)$, we may also obtain approximate expressions for the decay rate in the strong and weak coupling limits: 
\begin{enumerate}
   
    \item[(i)] Strong coupling limit $g\gg\Gamma$: using $\Omega = g\sqrt{1-(\Gamma/4g)^2} \approx g$, then 
    \begin{equation}\label{eq:gm_sc}
        \gamma(t) \approx \frac{2g^2\sin(g t)}{g\cos(g t) + \tfrac{\Gamma}{4}\sin(g t)} \approx 2g\tan(gt).
    \end{equation}
    
    \item[(ii)] Weak coupling limit $g\ll \Gamma$: using $\Omega = i\tfrac{\Gamma}{4}\sqrt{1 - (4g/\Gamma)^2} \approx i\Gamma/4$, we first have
    \begin{equation}\label{eq:gm_wc_1}
        \gamma(t) \approx \frac{2g^2\sin(\tfrac{i\Gamma t}{4})}{i\tfrac{\Gamma}{4}\cos(\tfrac{i\Gamma t}{4}) + \tfrac{\Gamma}{4}\sin(\tfrac{i\Gamma t}{4})} 
                  = \frac{8g^2}{\Gamma}\frac{\sinh(\tfrac{\Gamma t}{4})}{\cosh(\tfrac{\Gamma t}{4}) + \sinh(\tfrac{\Gamma t}{4})}. 
    \end{equation}
    The relevant time scale in this expression is $\tau_B\sim 1/\Gamma$, where $\tau_B$ denotes the bath correlation time \cite{Gardiner2009} (i.e., the time scale on which the bath correlation function \eqref{eq:bcf_lorz} decays in the long time limit).
    On the other hand, the relaxation time scale of the system is $\tau_R \sim 1/\gamma_0$, where $\gamma_0=2{\rm Re}\int^{\infty}_0dt\,f(t) = 4g^2/\Gamma$ \cite{BP2002}. 
    Hence, for times $t\sim \tau_R\gg\tau_B$, the decay rate \eqref{eq:gm_wc_1} may be approximated by its asymptotic value
    \begin{equation}\label{eq:gm_wc_2}
        \gamma(t) \approx \lim_{t\rightarrow\infty}\frac{8g^2}{\Gamma}\frac{\sinh(\tfrac{\Gamma t}{4})}{\cosh(\tfrac{\Gamma t}{4}) + \sinh(\tfrac{\Gamma t}{4})} = \gamma_0,
    \end{equation}
    which equals the Markovian emission rate appearing in the master equation \eqref{eq:sb_gksl}. 
    In this respect, we note that the limit $\tau_R\gg\tau_B$ coincides with the Born-Markov approximations under which the Markovian master equation \eqref{eq:sb_gksl} can be microscopically derived \cite{BP2002} (compare also with the random walk example outlined in Sec. \ref{sec:2.1.1}).

\end{enumerate}

\begin{figure}[t!]
    \centering
    \begin{subfigure}{0.45\textwidth}
        \caption{}
        \includegraphics[scale=0.7]{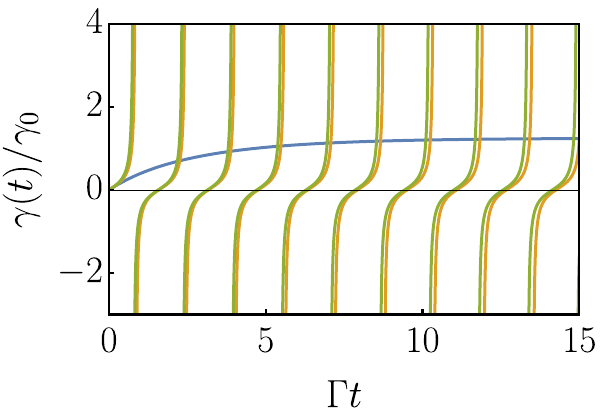}
    \end{subfigure}
    \begin{subfigure}{0.45\textwidth}
        \caption{}
        \includegraphics[scale=0.7]{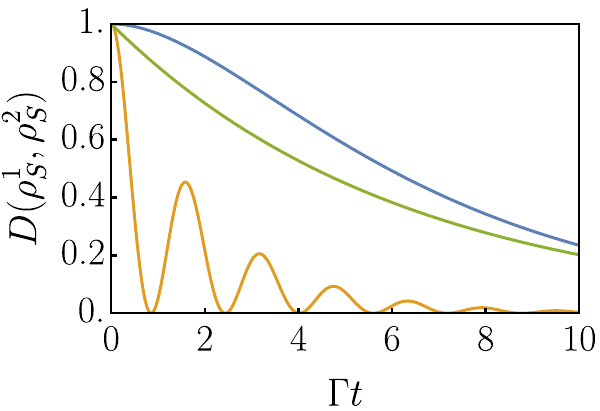}
    \end{subfigure}
    \caption{(a) Decay rate $\gamma(t)$ for parameters $g = 0.2\Gamma$ (blue) and $g = 2\Gamma$ (orange). 
    The blue line corresponds to the Markovian regime where the decay rate converges to $\gamma_0=4g^2/\Gamma$ in the long time limit $t\rightarrow\infty$; see \eqref{eq:gm_wc_2}.
    Conversely, the green line shows the approximate form \eqref{eq:gm_sc} derived in the strong coupling limit for $g=2\Gamma$. 
    (b) Trace distance for parameters $g=0.2\Gamma$ (blue), $g=2\Gamma$ (orange), and $\rho^1_S=|1\rangle\langle 1|$, $\rho^2_S=|0\rangle\langle0|$. 
    The green line shows \eqref{eq:td_markov} derived from the GKSL master equation \eqref{eq:sb_gksl} for $g=0.2\Gamma$, which closely matches the blue line at times $t\gg1/\Gamma$.}
    \label{fig:3}
\end{figure}

Figure \ref{fig:3} displays the behavior of the decay rate $\gamma(t)$ and trace distance for different values of the coupling strength $g$ with $\Delta\rho_{11}=1$ and $\Delta\rho_{10}=0$.
For moderately strong coupling $g=2\Gamma$, the system displays non-Markovian behavior since the decay rate is periodic and clearly takes negative values.  
This behavior is also reflected in the trace distance insofar that $\sigma(\rho^{1,2}_S,t)$ is positive during the same time intervals where $\gamma(t)<0$. 
For weak coupling $g=0.2\Gamma$, the decay rate is always positive and can be well-approximated by \eqref{eq:gm_wc_2} on the system time scale $\tau_R\sim 1/\gamma_0$.
In parallel, by substituting the decay rate $\gamma(t)\approx\gamma_0$ and propagator \eqref{eq:G_exact} into \eqref{eq:td_sb}, the trace distance in the weak coupling limit can be shown to approximately equal 
\begin{equation}\label{eq:td_markov}
    D(\rho^1_S(t),\rho^2_S(t)) \approx e^{-\gamma_0t}, \qquad t\gg1/\Gamma,
\end{equation}
which matches the prediction of the Markovian master equation \eqref{eq:sb_gksl}.

\section{Exercises}

\begin{enumerate}

    \item Prove that the joint probability distributions \eqref{eq:joint_prob} are both normalized and satisfy the Kolmogorov consistency conditions \eqref{eq:kcc1}-\eqref{eq:kcc2}. 

    \item For a density matrix $\rho$ expressed in the orthonormal basis $\{|i\rangle\}$, the transposition map $\Phi^T$ describes the operation \cite{Lidar2020,Vacchini2024}
    \begin{equation*}
        \Phi^T(\rho) = \sum_{i,j}\rho_{ij}\Phi^T(|i\rangle\langle j|) = \sum_{i,j}\rho_{ij}|j\rangle\langle i| = \rho^T.  
    \end{equation*}
    Prove that such a map is (i) positive, and (ii) non-CP. 
    
    \item Prove that both a necessary and sufficient condition for the map \eqref{eq:map_sb} to be CP is $\int^t_0ds\,\gamma(s) \geq 0$.  

    \item For the spin-boson example in Sec. \ref{sec:4}, show that the trace distance $D(\rho^1_S(t),\rho^2_S(t))$ and its time derivative $\sigma(\rho^{1,2}_S,t) = \tfrac{d}{dt}D(\rho^1_S(t),\rho^2_S(t))$ are given by \eqref{eq:td_sb} and \eqref{eq:td_sb_deriv}, respectively. 
            
    \item Consider the pure dephasing dynamics of a qubit described by the time-local master equation
    \begin{equation}\label{eq:pure_dephase}
        \frac{d}{dt}\rho_S(t) = \frac{\gamma(t)}{2}[\sigma_z\rho_S(t)\sigma_z - \rho_S(t)].
    \end{equation}
    Show using \eqref{eq:g_tl} that the corresponding RHP measure is \cite{Rivas2010b}
    \begin{equation*}
        \mathcal{N}_{\rm RHP}(\Phi) = 
        \begin{cases}
            0 & \text{if $\gamma(t)\geq 0$}, \\
            \int_{\gamma<0}dt\,|\gamma(t)| & \text{if $\gamma(t)< 0$}.
        \end{cases}
    \end{equation*}
    
\end{enumerate}

\appendix

\counterwithin{equation}{section} 
\renewcommand{\theequation}{\thesection\arabic{equation}}

\section{Microscopic derivation of the GKSL master equation}\label{appenA}

In this section we outline key parts of the microscopic derivation of the GKSL master equation \eqref{eq:gksl}, starting from a generic Hamiltonian describing an open quantum system $S$ in contact with a heat bath $B$ (macroscopic environment in thermal equilibrium). Since this derivation is comprehensively treated in several textbooks \cite{BP2002,Rivas2010,Vacchini2024,Lidar2020}, we mainly focus here on the physical approximations involved and omit much of the technical detail.

The composite Hamiltonian of an open quantum system and heat bath may generally be written in the form
\begin{equation}
    H = H_S + H_B + \lambda H_I,
\end{equation}
where the first two terms $H_S$ and $H_B$ describe the free evolution of $S$ and $B$ when they are uncoupled. The third term $H_I$ describes their mutual interaction and admits the decomposition
\begin{equation}\label{eq:H_int_appen}
    H_I = \sum_{\alpha}A_{\alpha}\otimes B_{\alpha},
\end{equation}
with $A_{\alpha}\in\mathcal{B}(\mathcal{H}_S)$ and $B_{\alpha}\in\mathcal{B}(\mathcal{H}_B)$ Hermitian operators of the system and bath, respectively.
The dimensionless parameter $\lambda$ tracks the order at which the system-bath coupling is evaluated within the evolution equation for the reduced system density matrix $\rho_S(t)$ (see below).  
We also remark that \eqref{eq:H_int_appen} defines a bilinear form of system-bath interaction where $B_{\alpha}$ is at most linear in operators of the bath (spin, bosonic, or fermionic), and it is assumed without loss of generality that ${\rm Tr}_B[B_{\alpha}\rho_B]=0$\footnote{As shown in \cite{Rivas2010,Lidar2020}, this can always be prearranged via the canonical transformation $B_{\alpha}\rightarrow B_{\alpha}-{\rm Tr}[B_{\alpha}\rho_B]$ and $H_S \rightarrow H_S + \sum_{\alpha}{\rm Tr}[B_{\alpha}\rho_B]A_{\alpha}$. }.

Given an initially factorized state $\rho(0)=\rho_S(0)\otimes\rho_B$ (see \eqref{eq:initial}), the von Neumann equation for $\rho(t) = U(t)\rho_S(0)\otimes \rho_BU^{\dagger}(t)$ in the interaction picture reads 
\begin{equation}
    \frac{d}{dt}\rho(t) = -i\lambda [H_I(t),\rho(t)],
\end{equation}
with $H_I(t) = e^{i(H_S+H_B)t}H_Ie^{-i(H_S+H_B)t}$. We now proceed by reinserting to formal solution to $\rho(t)$ back into the right hand side of the above and marginalizing over $B$ on both sides. After a change of variable, we obtain 
\begin{equation}\label{eq:appenA_1}
    \frac{d}{dt}\rho_S(t) = -\lambda^2\int^t_0ds\,{\rm Tr}_B[H_I(t),[H_I(t-s),\rho(t-s)]],
\end{equation}
where the linear term proportional to ${\rm Tr}_B[H_I(t),\rho(0)]$ vanishes since ${\rm Tr}_B[B_{\alpha}\rho_B]=0$. 

At this stage it is feasible to introduce two key approximations underpinning the physical validity of the GKSL master equation \eqref{eq:gksl}. The first of these is the so-called Born approximation, in which the system-bath density matrix is assumed for $t\geq 0$ to approximately factorize as 
\begin{equation}\label{eq:born}
    \rho(t)\approx\rho_S(t)\otimes \rho_B.
\end{equation}
More specifically, this amounts to assuming weak system-bath coupling insofar that the bath is only very weakly perturbed by its interaction with $S$, and hence remains in equilibrium on the timescale $\tau_R$ over which $\rho_S(t)$ appreciably evolves. It is important to note, however, that this approximation is only formally valid when used in the perturbative expansion of the generator \eqref{eq:tlg}, since under a global unitary evolution $S$ and $B$ will generally become correlated even if they are weakly coupled (this point is clearer in the alternative projection operator derivation \cite{BP2002,Vacchini2024,Lidar2020}). By further noting that 
\begin{equation}
    \rho_S(t-s) = \rho_S(t) + \int^{t-s}_tdu\frac{\partial}{\partial u}\rho_S(u) = \rho_S(t) + O(\lambda^2),
\end{equation}
then \eqref{eq:appenA_1} can be rewritten as 
\begin{equation}
    \frac{d}{dt}\rho_S(t) = -\lambda^2\int^t_0ds\,{\rm Tr}_B[H_I(t),[H_I(t-s),\rho_S(t)]] + O(\lambda^3). 
\end{equation}
If we now compare the first term with the exact equation $\frac{d}{dt}\rho_S(t)=\mathcal{L}_t\rho_S(t)$, it is evident that it represents the leading order contribution in the expansion $\mathcal{L}_t = \lambda^2\mathcal{L}^{(2)}_t + O(\lambda^3)$. Since terms $O(\lambda^3)$ are negligible under the Born approximation, we may therefore simplify the above to 
\begin{equation}\label{eq:appenA_2}
    \frac{d}{dt}\rho_S(t) \approx -\lambda^2\sum_{\alpha,\beta}\int^t_0ds\Big(C_{\alpha\beta}(t-s)[A_{\alpha}(t),A_{\beta}(t-s)\rho_S(t)] + {\rm h.c.}\Big),
\end{equation}
where we have defined $A_{\alpha}(t) = e^{iH_St}A_{\alpha}e^{-iH_St}$ and introduced the bath correlation functions
\begin{equation}\label{eq:bcf}
    C_{\alpha\beta}(t-s) = {\rm Tr}[B_{\alpha}(t)B_{\beta}(s)\rho_B],
\end{equation}
with $B_{\alpha}(t) = e^{iH_Bt}B_{\alpha}e^{-iH_Bt}$. Note that for these functions to be time homogeneous it is sufficient for the initial bath state to satisfy $[H_B,\rho_B]=0$, which is indeed the case for a thermal state $\rho_B = e^{-H_B/k_BT}/{\rm Tr}[e^{-H_B/k_BT}]$ parameterized by the bath temperature $T$. 

The second key approximation we make relates to removing the effect of the bath memory on the reduced system dynamics, known as the Markovian approximation. In particular, if the bath correlation functions decay asymptotically as $C_{\alpha\beta}(t-s)\sim e^{-|t-s|/\tau_B}$, where $\tau_B$ is the bath correlation time, then the Markovian approximation is equivalent to assuming a separation of timescales
\begin{equation}\label{eq:appen_markov}
    \tau_R\gg \tau_B.
\end{equation}
Note this is very similar to the memoryless approximation discussed for the colloidal particle example in Sec. \ref{sec:2}.
A corollary of this assumption is that the upper limit of integral in \eqref{eq:appenA_2} may be extended to infinity while only negligibly affecting the result: 
\begin{equation}
    \frac{d}{dt}\rho_S(t) = -\lambda^2\sum_{\alpha,\beta}\int^{\infty}_0ds\Big(C_{\alpha\beta}(t-s)[A_{\alpha}(t),A_{\beta}(t-s)\rho_S(t)] + {\rm h.c.}\Big).
\end{equation}
To further simplify this equation, we employ the following spectral decomposition of the system operators $A_{\alpha}(t)$,
\begin{equation}
    A_{\alpha}(t) = \sum_{\omega}A_{\alpha}(\omega)e^{-i\omega t}.
\end{equation}
Here, $A_{\alpha}(\omega)$ are eigenoperators of the system Hamiltonian $[H_S,A_{\alpha}(\omega)] = -\omega A_{\alpha}(\omega)$ with the corresponding eigenfrequencies $\omega = \epsilon - \epsilon'$, and $H_S|\epsilon\rangle = \epsilon|\epsilon\rangle$ (provided $H_S$ has a discrete spectrum) \cite{BP2002}. Since $A_{\alpha}(t)$ is Hermitian we have the relation
\begin{equation}
    \sum_{\omega}A_{\alpha}(\omega)e^{-i\omega t} = \sum_{\omega}A^{\dagger}_{\alpha}(\omega)e^{i\omega t},
\end{equation}
such that the above can be expressed as 
\begin{equation}\label{eq:appenA_3}
    \frac{d}{dt}\rho_S(t) = -\lambda^2\sum_{\alpha,\beta}\sum_{\omega,\omega'}\Big(\Gamma_{\alpha\beta}(\omega)e^{i(\omega'-\omega)t}[A^{\dagger}_{\alpha}(\omega'),A_{\beta}(\omega)\rho_S(t)] + {\rm h.c.}\Big),
\end{equation}
with $\Gamma_{\alpha\beta}(\omega) = \int^{\infty}_0dt\,C_{\alpha\beta}(t)e^{i\omega t}$.   

We are now in a position to make the final approximation needed to put the master equation into GKSL form---the secular approximation---which, similar to \eqref{eq:appen_markov}, relies on a separation of timescales argument to neglect terms from \eqref{eq:appenA_3} that oscillate rapidly over the evolution timescale $\tau_R$. To do so, we first introduce the system timescale $\tau_S = 1/|\omega - \omega'|$, such that $\tau^{-1}_S$ measures the rate at which $S$ evolves according to $H_S$. The secular approximation then neglects the $\omega\neq\omega'$ terms under the assumption
\begin{equation}
    \tau_R \gg \tau_S.
\end{equation}
We remark that in some literature this is also called the rotating-wave approximation \cite{Lidar2020}. Lastly, we decompose the matrix $\Gamma_{\alpha\beta}(\omega)$ into its Hermitian and anti-Hermitian parts 
\begin{equation}
    \Gamma_{\alpha\beta}(\omega) = \frac{1}{2}\gamma_{\alpha\beta}(\omega) + iS_{\alpha\beta}(\omega),
\end{equation}
with  
\begin{equation}
    S_{\alpha\beta}(\omega) = \frac{1}{2i}[\Gamma_{\alpha\beta}(\omega) - \Gamma^*_{\beta\alpha}(\omega)], \qquad \gamma_{\alpha\beta}(\omega) = \Gamma_{\alpha\beta}(\omega) + \Gamma^*_{\beta\alpha}(\omega),
\end{equation}
which in \eqref{eq:appenA_3} yields (reabsorbing $\lambda$ back into the bath operators $B_{\alpha}$) 
\begin{align}\label{eq:gksl_me}
    \frac{d}{dt}\rho_S(t) = \mathcal{L}\rho_S(t) &= -i\Big[\sum_{\alpha,\beta}\sum_{\omega}S_{\alpha\beta}(\omega)A^{\dagger}_{\alpha}(\omega)A_{\beta}(\omega), \rho_S(t)  \Big] \nonumber\\
    &+ \sum_{\alpha,\beta}\sum_{\omega}\gamma_{\alpha\beta}(\omega)\Big[A_{\beta}(\omega)\rho_S(t)A^{\dagger}_{\alpha}(\omega) - \tfrac{1}{2}\{A^{\dagger}_{\alpha}(\omega)A_{\beta}(\omega),\rho_S(t)\}\Big]. 
\end{align}
Diagonalizing $\boldsymbol{\gamma}(\omega)$ then reveals that the generator $\mathcal{L}$ admits the same GKSL form as \eqref{eq:gksl}.

In summary, master equations of the type \eqref{eq:gksl_me} may be physically applied in situations where the following three approximations are valid: 
\begin{itemize}
    \item[(i)] Born approximation: the system-bath coupling is sufficiently weak that terms beyond second-order in $\lambda$ may be neglected from $\mathcal{L}_t = \lambda^2\mathcal{L}^{(2)}_t+O(\lambda^3)$.
    \item[(ii)] Markov approximation: the bath correlation time $\tau_B$ is sufficiently short that integrals containing the memory kernels $C_{\alpha\beta}(t-s)$ may be extended to infinity.
    \item[(iii)] Secular approximation: the free evolution timescale $\tau_S$ is sufficiently short that nonsecular terms $\sim e^{i(\omega'-\omega)t}$ ($\omega\neq\omega'$) may be neglected from \eqref{eq:appenA_3}. 
\end{itemize}

\section{Proof of Lemma 6}\label{appenB} 

We start by proving the forward statement CPTP $\Lambda$ $\Rightarrow$ $\|C_{\Lambda}\|_1 = 1$. 
Since $C_{\Lambda}$ is Hermitian and nonnegative, it admits a spectral decomposition 
$C_{\Lambda} = \sum^{d^2}_{i=1}\lambda_i|\lambda_i\rangle\langle\lambda_i|$ with $\lambda_i\geq 0$:
thus the trace norm of $C_{\Lambda}$ reads 
\begin{align}
   \|C_{\Lambda}\|_1 := {\rm Tr}\sqrt{C^{\dagger}_{\Lambda}C_{\Lambda}} = \sum^d_{i=1}|\lambda_i| = \sum^d_{i=1}\lambda_i.
\end{align}
The same Choi matrix also admits the trace 
\begin{equation}\label{eq:C_tr}
    {\rm Tr}\,C_{\Lambda} = \sum^d_{i=1} \lambda_i = \|C_{\Lambda}\|_1. 
\end{equation}
Furthermore, 
\begin{align}\label{eq:appen1}
    {\rm Tr}\,C_{\Lambda} &= \frac{1}{d}{\rm Tr}\bigg[\sum_{i,j}\Lambda(|i\rangle\langle j|) \otimes |i\rangle\langle j|\bigg] \nonumber\\
                          &= \frac{1}{d}\sum_{i,j}{\rm Tr}[\Lambda(|i\rangle\langle j|)] \otimes \underbrace{{\rm Tr}(|i\rangle\langle j|)}_{=\delta_{ij}} \nonumber\\
                          &= \frac{1}{d}\sum_{i,j}{\rm Tr}[\Lambda(|i\rangle\langle i|)] && \hspace{-3cm} {\rm (Trace \,\, preserving)} \nonumber\\
                          &= \frac{1}{d}\sum^d_{i=1}{\rm Tr}[|i\rangle\langle i|] = 1,
\end{align}
such that $\|C_{\Lambda}\|_1 = 1$. 

Next, we prove the converse statement $\|C_{\Lambda}\|_1 = 1$ $\Rightarrow$ CPTP $\Lambda$. 
For this we assume the map $\Lambda$ to be trace and Hermiticity preserving.
As such, we have from the spectral decomposition of $C_{\Lambda}$ that 
\begin{equation}
    C_{\Lambda} = C^+_{\Lambda} - C^-_{\Lambda}, 
\end{equation}
where $C^+_{\Lambda}=\sum_i\lambda^+_i|\lambda^+_i\rangle\langle\lambda^+_i|$ ($C^-_{\Lambda}=\sum_i\lambda^-_i|\lambda^-_i\rangle\langle\lambda^-_i|$) is a positive operator containing the positive (negative) eigenvalues of $C_{\Lambda}$.
It is also the case from \eqref{eq:appen1} that   
\begin{equation}\label{eq:appen2}
    {\rm Tr}\,C_{\Lambda} = {\rm Tr}\,C^+_{\Lambda} - {\rm Tr}\,C^-_{\Lambda} = 1,
\end{equation}
and from $C^{\pm}_{\Lambda}\geq 0$,
\begin{align}\label{eq:appen3}
    \|C_{\Lambda}\|_1 &= \sum_i|\lambda^+_i| + \sum_i|\lambda^-_i| \nonumber\\
                      &={\rm Tr}\,C^+_{\Lambda} + {\rm Tr}\,C^-_{\Lambda} = 1 \geq {\rm Tr}\,C^+_{\Lambda} - {\rm Tr}\,C^-_{\Lambda}.
\end{align}
Hence, via \eqref{eq:C_tr}, \eqref{eq:appen2} and \eqref{eq:appen3} we must have ${\rm Tr}\,C^-_{\Lambda} = 0$, which implies $C^-_{\Lambda} = 0$ and $C_{\Lambda} = C^+_{\Lambda}$.
Since $C_{\Lambda}$ is then guaranteed to be nonnegative, this tells us $\Lambda$ is CP according to Choi's theorem,
and so $\Lambda$ CPTP $\Leftrightarrow$ $\|C_{\Lambda}\|_1 = 1$. Alternatively, if $\Lambda$ is non-CP yet both Hermiticity and trace preserving,  
\eqref{eq:appen3} implies $\|C_{\Lambda}\|_1 = {\rm Tr}\,C^+_{\Lambda} + {\rm Tr}\,C^-_{\Lambda} > 1$. \qed 

\bigskip 

\section*{Solutions to Exercises}

\subsection*{Exercise 1}

It is easily demonstrated that \eqref{eq:joint_prob} has unit normalization through $\sum_iT_{ij}=1$:
\begin{align*}
    \sum_{x_{t_n},...,x_{t_1}}P_n(x_{t_n}, ..., x_{t_1}) &= \sum_{x_{t_n},...,x_{t_1}}T_{x_{t_n}x_{t_{n-1}}}...\,T_{x_{t_2}x_{t_1}}p_{x_{t_1}}  \\
                                                         &= \underbrace{\bigg(\sum_{x_{t_n}}T_{x_{t_n}x_{t_{n-1}}}\bigg)}_{=1}...\underbrace{\bigg(\sum_{x_{t_2}}T_{x_{t_2}x_{t_1}}\bigg)}_{=1}\sum_{x_{t_1}}p_{x_{t_1}} \\
                                                         &= \sum_{x_{t_1}}p_{x_{t_1}} = 1.
\end{align*}
We can now verify the first of the Kolmogorov consistency relations by marginalizing the distribution over $x_{t_j}$. 
To do so, we consider three separate cases:
\begin{align*}
   \hspace{-1cm}  j=1:  \hspace{1cm} \sum_{x_{t_1}}P_n(x_{t_n},...,x_{t_1}) &= T_{x_{t_n}x_{t_{n-1}}}...\sum_{x_{t_1}}T_{x_{t_2}x_{t_1}}p_{x_{t_1}}
                                                                           = T_{x_{t_n}x_{t_{n-1}}}... \,T_{x_{t_3}x_{t_2}}p_{x_{t_2}} \\
                                                                          &= P_{n-1}(x_{t_n},...,x_{t_2}). \\
    \hspace{-1cm} n>j>1:    \hspace{1cm}  \sum_{x_{t_j}}P_n(x_{t_n},...,x_{t_1}) &= T_{x_{t_n}x_{t_{n-1}}}...\sum_{x_{t_j}}T_{x_{t_{j+1}}x_{t_j}}T_{x_{t_j}x_{t_{j-1}}}...\,T_{x_{t_2}x_{t_1}}p_{x_{t_1}} \\
                                                                               &= T_{x_{t_n}x_{t_{n-1}}}...\, T_{x_{t_{j+1}}x_{t_{j-1}}}...\,T_{x_{t_2}x_{t_1}}p_{x_{t_1}} \hspace{1cm} \text{(CKE)}  \\
                                                                               &= P_{n-1}(x_{t_n},...x_{t_{j+1}},x_{t_{j-1}},...,x_{t_1}). \\
     \hspace{-1cm} j=n:    \hspace{1cm}  \sum_{x_{t_n}}P_n(x_{t_n},...,x_{t_1}) &= \sum_{x_{t_n}}T_{x_{t_n}x_{t_{n-1}}}...\, T_{x_{t_2}x_{t_1}}p_{x_{t_1}}
                                                                              = T_{x_{t_{n-1}x_{t_{n-2}}}}... \,T_{x_{t_2}x_{t_1}}p_{x_{t_1}} \\ 
                                                                              &= P_{n-1}(x_{t_{n-1}},...,x_{t_1}).                                                                                         
\end{align*}
Note these results hold regardless of the ordering of the indices $x_{t_j}$. Hence, \eqref{eq:joint_prob} satisfies the Kolmogorov consistency relations.

\subsection*{Exercise 2}

We follow the approach of Example \ref{example:2} and start by proving the positivity of $\Phi^T$.
Since $X\geq 0$ implies $X=\sum_i\lambda_i|i\rangle\langle i|$, where $\lambda_i$ are real and positive, then
\begin{equation*}
    \Phi^T(X) = \sum_i\lambda_i\Phi^T(|i\rangle\langle i|) = \sum_i\lambda_i|i\rangle\langle i| \geq 0.
\end{equation*}
A counterexample to $(\Phi^T\otimes\mathcal{I}^{(d)}_A)|\Phi^+_d\rangle\langle\Phi^+_d|$ being positive can be realized for $d=2$. 
In particular, 
\begin{align*}
    \frac{1}{2}\sum^2_{i,j=1}\Phi^T(|i\rangle\langle j|)\otimes|i\rangle\langle j| = \frac{1}{2}\sum^2_{i,j=1}|j\rangle\langle i|\otimes|i\rangle\langle j|,
\end{align*}
whose corresponding eigenvalues read $\{1/2,1/2,1/2,-1/2\}$. Hence, the transposition map $\Phi^T$ is non-CP.

\subsection*{Exercise 3}

The conditions under which the map \eqref{eq:map_sb} is CP can be established through examining its Choi matrix: 
\begin{equation*}
    C_{\Phi}(t) = \frac{1}{2}\sum^1_{i,j=0}\Phi_t(|i\rangle\langle j|)\otimes|i\rangle\langle j|. 
\end{equation*} 
To proceed, we need to deduce the image of $|i\rangle\langle j|$ under $\Phi_t$.
It is evident from \eqref{eq:G_exact} that $\Phi_t$ admits the steady state $|0\rangle\langle 0|$, 
implying this component is time invariant,
while all remaining components must vanish in the limit $t\rightarrow\infty$.  
Likewise, the image of $|1\rangle\langle 1|$ should contain $|G(t)|^2|1\rangle\langle 1|$ (see \eqref{eq:map_sb}),
as well as an additional term $\sim |0\rangle\langle 0|$ needed to maintain the correct normalization of $\rho_S(t)$. 
Taking all this into account, we have \cite{Lidar2020}
\begin{align*}
    \Phi_t(|1\rangle\langle 1|) &= |G(t)|^2|1\rangle\langle 1| + (1-|G(t)|^2)|0\rangle\langle 0| \\
    \Phi_t(|0\rangle\langle 0|) &= |0\rangle\langle 0| \\
    \Phi_t(|1\rangle\langle 0|) &= G(t)|1\rangle \langle 0| \\
    \Phi_t(|0\rangle\langle 1|) &= G^*(t)|0\rangle\langle 1|. 
\end{align*}
Hence, the Choi matrix reads 
\begin{align*}
    C_{\Phi}(t) &= \frac{1}{2}\Big[|G(t)|^2|1\rangle\langle 1|\otimes|1\rangle\langle 1| + (1 - |G(t)|^2)|0\rangle\langle 0|\otimes|1\rangle\langle 1|   \\
                &\qquad + |0\rangle\langle 0|\otimes|0\rangle\langle 0| + G(t)|1\rangle\langle 0|\otimes|1\rangle\langle 0| + G^*(t)|0\rangle\langle 1|\otimes |0\rangle\langle 1|\Big],
\end{align*}
which can be written in the ordered basis $\{|11\rangle,|10\rangle,|01\rangle,|00\rangle\}$ as 
\begin{equation*}
    C_{\Phi}(t) = \frac{1}{2}
    \begin{pmatrix}
        |G(t)|^2 & 0 & 0 & G(t) \\
        0 & 0 & 0 & 0 \\
        0 & 0 & 1 - |G(t)|^2 & 0 \\
        G^*(t) & 0 & 0 & 1
    \end{pmatrix}.
\end{equation*}
By inspection, this matrix must have at least one zero eigenvalue since its second row maps any nonzero input $\boldsymbol{v} = (v_1,v_2,v_3,v_4)^T$ to zero, i.e. $[C_{\Phi}(t)\boldsymbol{v}]_2 = 0\cdot v_1 + 0\cdot v_2 + 0\cdot v_3 + 0\cdot v_4$. 
By the same token, another eigenvalue must equal $1-|G(t)|^2$. 
The remaining two eigenvalues can be found by diagonalizing the $2\times 2$ matrix
\begin{equation*}
    \frac{1}{2}
    \begin{pmatrix}
        |G(t)|^2 & G(t) \\
        G^*(t) & 1
    \end{pmatrix},
\end{equation*}
which is easily done to obtain 
\begin{equation*}
    {\rm spec}\,C_{\Phi}(t) = \big\{0, \tfrac{1}{2}(1-|G(t)|^2), 0, \tfrac{1}{2}(1+|G(t)|^2)\big\}. 
\end{equation*}
We now require these eigenvalues to be strictly nonnegative for $\Phi_t$ to be CP. 
The only eigenvalue which may become negative is $\tfrac{1}{2}(1-|G(t)|^2)$.
Thus, from \eqref{eq:G_exact} we require $\exp(-\int^t_0ds\,\gamma(s))\leq 1$, so that via Theorem \ref{theorem:3},
\begin{equation*}
    \int^t_0ds\,\gamma(s) \geq 0, \qquad t\geq 0
\end{equation*}
provides a necessary and sufficient condition for $\Phi_t$ to be CP.

\subsection*{Exercise 4}

The matrix $\Delta\rho_S(t) = \rho^1_S(t) - \rho^2_S(t)$ can be written in the basis $\{|1\rangle,|0\rangle\}$ as 
\begin{equation*}
    \Delta \rho_S(t) := 
        \begin{pmatrix}
        \Delta\rho_{11}(t) & \Delta\rho_{10}(t) \\
        \Delta\rho_{10}^*(t) & \Delta\rho_{00}(t)
    \end{pmatrix}
    =
    \begin{pmatrix}
        |G(t)|^2\Delta\rho_{11} & G(t)\Delta\rho_{10} \\
        G^*(t)\Delta\rho_{11}^* & -|G(t)|^2\Delta\rho_{11}
    \end{pmatrix}.
\end{equation*}
Since its trace norm is simply the sum of its eigenvalues, $\|\Delta\rho_S(t)\|_1=|\lambda_+|+|\lambda_-|$, we proceed by solving the characteristic equation
\begin{align*}
    |\lambda\mathbb{I} - \Delta\rho_S(t)| = \lambda^2 - [\Delta\rho_{11}(t) + \Delta\rho_{00}(t)]\lambda + \Delta\rho_{11}(t)\Delta\rho_{00}(t) - |\Delta\rho_{10}(t)|^2 = 0, 
\end{align*}
where
\begin{align*}
    \lambda_{\pm} &= \tfrac{1}{2}\Big(\Delta\rho_{11}(t) + \Delta\rho_{00}(t) \pm \sqrt{\Delta\rho^2_{11}(t) + \Delta\rho^2_{00}(t) - 2\Delta\rho_{11}(t)\Delta\rho_{00}(t) + 4|\Delta\rho_{10}(t)|^2}\Big) \\
                  &= \tfrac{1}{2}\Big(\Delta\rho_{11}(t) + \Delta\rho_{00}(t) \pm \sqrt{[\Delta\rho_{11}(t) - \Delta\rho_{00}(t)]^2 + 4|\Delta\rho_{10}(t)|^2}\Big).
\end{align*}
Now using 
\begin{align*}
    \Delta\rho_{11}(t) + \Delta\rho_{00}(t) = 0, \quad \Delta\rho_{11}(t) - \Delta\rho_{00}(t) = 2|G(t)|^2\Delta\rho_{11}, \quad |\Delta\rho_{10}(t)|^2 = |G(t)|^2|\Delta\rho_{10}|^2,
\end{align*}
we get 
\begin{equation*}
    \lambda_{\pm} = \pm |G(t)|\sqrt{|G(t)|^2\Delta\rho^2_{11} + |\Delta\rho_{10}|^2},
\end{equation*}
such that 
\begin{equation*}
    \frac{1}{2}\|\Delta\rho_S(t)\|_1 = |G(t)|\sqrt{|G(t)|^2\Delta\rho^2_{11} + |\Delta\rho_{10}|^2}.
\end{equation*}

\subsection*{Exercise 5}

To evaluate the RHP measure for the pure dephasing master equation \eqref{eq:pure_dephase}, we start from the expression
\begin{equation*}
    g(t) = \lim_{\epsilon\rightarrow0^+}\frac{\|(\mathcal{I} + \epsilon\mathcal{L}_t\otimes\mathcal{I}^{(2)}_A)|\Phi^+_2\rangle\langle\Phi^+_2|\|_1 - 1}{\epsilon}
\end{equation*}
with the time-local generator
\begin{equation*}
    \mathcal{L}_t\rho = \frac{\gamma(t)}{2}[\sigma_z\rho\sigma_z - \rho].
\end{equation*}
We know from Sec. \ref{sec:RHP} that $|\Phi^+_2\rangle\langle\Phi^+_2|$ can be expanded in the orthonormal basis $\{\tfrac{1}{\sqrt{2}}\mathbb{I}_S, \sigma_+, \sigma_-, \tfrac{1}{\sqrt{2}}\sigma_z\}$ as 
\begin{equation*}
    |\Phi^+_2\rangle\langle\Phi^+_2| = \tfrac{1}{4}\mathbb{I}_S\otimes\mathbb{I}_S + \tfrac{1}{2}\sigma_+\otimes\sigma_+ + \tfrac{1}{2}\sigma_-\otimes\sigma_- + \tfrac{1}{4}\sigma_z\otimes\sigma_z.
\end{equation*}
In turn, this expansion may be used evaluate $(\mathcal{L}_t\otimes\mathcal{I}^{(2)}_A)|\Phi^+_2\rangle\langle\Phi^+_2| = \sum_ic_{ii}\mathcal{L}_t[\sigma_i]\otimes\sigma_i$ through the following identities
\begin{align*}
    \mathcal{L}_t[\sigma_0] &= 0  \\
    \mathcal{L}_t[\sigma_1] &= -\gamma(t)\sigma_1 \\
    \mathcal{L}_t[\sigma_2] &= -\gamma(t)\sigma_2 \\
    \mathcal{L}_t[\sigma_3] &= 0, 
\end{align*} 
such that 
\begin{equation*}
    (\mathcal{I} + \epsilon\mathcal{L}_t\otimes\mathcal{I}^{(2)}_A)|\Phi^+_2\rangle\langle\Phi^+_2| = \tfrac{1}{4}\mathbb{I}_S\otimes\mathbb{I}_S + \tfrac{1}{2}(1-\gamma(t)\epsilon)\sigma_+\otimes\sigma_+ + \tfrac{1}{2}(1-\gamma(t)\epsilon)\sigma_-\otimes\sigma_- + \tfrac{1}{4}\sigma_z\otimes\sigma_z.
\end{equation*}
The Choi matrix written in the basis $\{|11\rangle,|10\rangle,|01\rangle,|00\rangle\}$ reads 
\begin{equation*}
    C_{\Lambda}(t+\epsilon,t) = (\mathcal{I} + \epsilon\mathcal{L}_t\otimes\mathcal{I}^{(2)}_A)|\Phi^+_2\rangle\langle\Phi^+_2| = 
    \frac{1}{2}
    \begin{pmatrix}
        1 & 0 & 0 & 1-\gamma(t)\epsilon \\
        0 & 0 & 0 & 0 \\
        0 & 0 & 0 & 0 \\
        1 - \gamma(t)\epsilon & 0 & 0 & 1
    \end{pmatrix},
\end{equation*}
whose trace norm $\|C_{\Lambda}(t+\epsilon,t)\|_1$ can be evaluated from its eigenvalues. Using the block structure of $C_{\Lambda}(t+\epsilon,t)$, we can immediately infer that at least two of its eigenvalues are zero, $\lambda_1 = \lambda_2 = 0$.
The remaining two eigenvalues are found by diagonalizing     
\begin{equation*}
    \begin{pmatrix}
        1 & 1 - \gamma(t)\epsilon \\ 
        1 - \gamma(t)\epsilon & 1
    \end{pmatrix}
\end{equation*}
which yields 
\begin{equation*}
    {\rm spec}\,C_{\Lambda}(t+\epsilon,t) = \big\{0, 0, 1 - \tfrac{1}{2}\gamma(t)\epsilon, \tfrac{1}{2}\gamma(t)\epsilon\big\}.
\end{equation*}
Accordingly, the trace norm is given by $\|C_{\Lambda}(t+\epsilon,t)\|_1 = |1 + \tfrac{1}{2}\gamma(t)\epsilon| + \tfrac{1}{2}|\gamma(t)|\epsilon$.
This can be expanded around $\epsilon=0$ to obtain 
\begin{equation*}
    \|C_{\Lambda}(t+\epsilon,t)\|_1 = 1 - \tfrac{1}{2}\gamma(t)\epsilon + \tfrac{1}{2}|\gamma(t)|\epsilon + O(\epsilon^2),
\end{equation*}
so that
\begin{equation*}
    g(t) = \tfrac{1}{2}[|\gamma(t)| - \gamma(t)]. 
\end{equation*}
Hence, the RHP measure for the pure dephasing master equation \eqref{eq:pure_dephase} is 
    \begin{equation*}
        \mathcal{N}_{\rm RHP}(\Phi) = 
        \begin{cases}
            0 & \text{if $\gamma(t)\geq 0$}, \\
            \int_{\gamma<0}dt\,|\gamma(t)| & \text{if $\gamma(t)< 0$}.
        \end{cases}
    \end{equation*}

\bigskip

\nolinenumbers

\textbf{Acknowledgements.} GP gratefully acknowledges the support of the National Institute for Theoretical and Computational Sciences (NITheCS) of South Africa. 
The author declares there are no competing interests. This manuscript does not report data.

\end{document}